\documentclass[aps,prl,reprint,superscriptaddress,showpacs,longbibliography,footnoteinbib]{revtex4-1}
\usepackage[latin9]{inputenc}
\usepackage{color}
\usepackage{bm}
\usepackage{amsmath}
\usepackage{amssymb}
\usepackage{graphicx}
\usepackage[unicode=true,
 bookmarks=false,
 breaklinks=false,pdfborder={0 0 1},backref=false,colorlinks=true]
 {hyperref}
\hypersetup{pdftitle={Title},
 plainpages=false,pdfpagelabels,linkcolor=blue,urlcolor=blue,citecolor=blue,pdfdisplaydoctitle=true,pdfduplex=DuplexFlipLongEdge}

\makeatletter
\usepackage{units}\usepackage{wasysym}

\definecolor{orange}{rgb}{0.50, 0.20, 0.0}

\newcommand{\beginsupplement}{%
	\setcounter{page}{1}
	 \renewcommand{\thepage}{SM - \arabic{page}}%
        \setcounter{table}{0}
        \renewcommand{\thetable}{S\arabic{table}}%
        \setcounter{figure}{0}
        \renewcommand{\thefigure}{S\arabic{figure}}%
        \setcounter{section}{0}
        \renewcommand{\thesection}{S\arabic{section}}%
        \setcounter{section}{0}
        \renewcommand{\thesection}{S\arabic{section}}%
        \setcounter{subsection}{0}
        \renewcommand{\thesubsection}{S\arabic{section}.\arabic{subsection}}%
        \setcounter{equation}{0}
        \renewcommand{\theequation}{S\arabic{equation}}%

     }

\usepackage{bm}
\usepackage{braket}

\makeatother

\begin{document}
\noindent\begin{minipage}[t]{1\columnwidth}%
\global\long\def\ket#1{\left| #1\right\rangle }%

\global\long\def\bra#1{\left\langle #1 \right|}%

\global\long\def\kket#1{\left\Vert #1\right\rangle }%

\global\long\def\bbra#1{\left\langle #1\right\Vert }%

\global\long\def\braket#1#2{\left\langle #1\right. \left| #2 \right\rangle }%

\global\long\def\bbrakket#1#2{\left\langle #1\right. \left\Vert #2\right\rangle }%

\global\long\def\av#1{\left\langle #1 \right\rangle }%

\global\long\def\tr{\text{tr}}%

\global\long\def\Tr{\text{Tr}}%

\global\long\def\pd{\partial}%

\global\long\def\im{\text{Im}}%

\global\long\def\re{\text{Re}}%

\global\long\def\sgn{\text{sgn}}%

\global\long\def\Det{\text{Det}}%

\global\long\def\abs#1{\left|#1\right|}%

\global\long\def\up{\uparrow}%

\global\long\def\down{\downarrow}%

\global\long\def\vc#1{\mathbf{#1}}%

\global\long\def\bs#1{\boldsymbol{#1}}%

\global\long\def\t#1{\text{#1}}%
\end{minipage}
\title{Renormalization-Group Theory of 1D quasiperiodic lattice models with
commensurate approximants}
\author{Miguel Gonçalves}
\affiliation{CeFEMA-LaPMET, Departamento de Física, Instituto Superior Técnico, Universidade de Lisboa, Avenida Rovisco Pais, 1049-001 Lisboa, Portugal}
\author{B.Amorim}
\affiliation{Centro de Física das Universidades do Minho e Porto (CF-UM-UP) and
Laboratory of Physics for Materials and Emergent Technologies LaPMET,
Universidade do Minho, Campus de Gualtar, 4710-057, Braga, Portugal}
\author{Eduardo V. Castro}
\affiliation{Centro de Física das Universidades do Minho e Porto, LaPMET, Departamento
de Física e Astronomia, Faculdade de Ciências, Universidade do Porto,
4169-007 Porto, Portugal}
\affiliation{Beijing Computational Science Research Center, Beijing 100193, China}
\author{Pedro Ribeiro}
\affiliation{CeFEMA, Instituto Superior Técnico, Universidade de Lisboa, Av. Rovisco
Pais, 1049-001 Lisboa, Portugal}
\affiliation{Beijing Computational Science Research Center, Beijing 100193, China}
\begin{abstract}
We develop a renormalization group (RG) description of the localization
properties of one-dimensional (1D) quasiperiodic lattice models. The
RG flow is induced by increasing the unit cell of subsequent commensurate
approximants. Phases of quasiperiodic systems are characterized by
RG fixed points associated with renormalized single-band models. We
identify fixed-points that include many previously reported exactly
solvable quasiperiodic models. By classifying relevant and irrelevant
perturbations, we show that phase boundaries of more generic models
can be determined with exponential accuracy in the approximant's unit
cell size, and in some cases analytically. Our findings provide a
unified understanding of widely different classes of 1D quasiperiodic
systems.
\end{abstract}
\maketitle
Quasiperiodic systems (QPS) offer a rich playground of exotic physics
ranging from interesting localization properties, in one \citep{AubryAndre,wilkinson,Roati2008,Lahini2009,Schreiber842,Luschen2018,slager1,PhysRevB.105.L220201}
or higher \citep{Huang2016a,PhysRevLett.120.207604,Park2018,PhysRevB.100.144202,Fu2020,Wang2020,Goncalves_2021}
dimensions, to topological non-trivial phases \citep{Kraus2012,PhysRevLett.109.116404,Verbin2013,slager2}.
QPS are of relevance to different platforms, including optical \citep{PhysRevA.75.063404,Roati2008,Modugno_2009,Schreiber842,Luschen2018,PhysRevLett.123.070405,PhysRevLett.125.060401,PhysRevLett.126.110401,PhysRevLett.122.170403}
and photonic lattices \citep{Lahini2009,Kraus2012,Verbin2013,PhysRevB.91.064201,Wang2020},
cavity-polariton devices \citep{Goblot2020Nature} and electronic
moiré systems \citep{Balents2020}.

The phase diagrams of one-dimensional (1D) QPS include extended, localized,
and even critical phases \citep{Liu2015,PhysRevLett.123.025301,anomScipost,PhysRevLett.125.073204}.
For special fine-tuned models, the phase diagram can be obtained analytically
\citep{AubryAndre,PhysRevB.43.13468,PhysRevLett.104.070601,PhysRevLett.113.236403,Liu2015,PhysRevB.91.235134,PhysRevLett.114.146601,Wang2020,anomScipost}.
However, generic phase diagrams are only numerically accessible for
finite systems \citep{PhysRevB.83.075105,Gopalakrishnan2017,Li2017,Szabo2018,PhysRevB.99.054211,PhysRevB.101.064203,Liu2017,PhysRevLett.126.106803,2021arXiv210103465L}.

\begin{figure}[h]
\begin{centering}
\includegraphics[width=1\columnwidth]{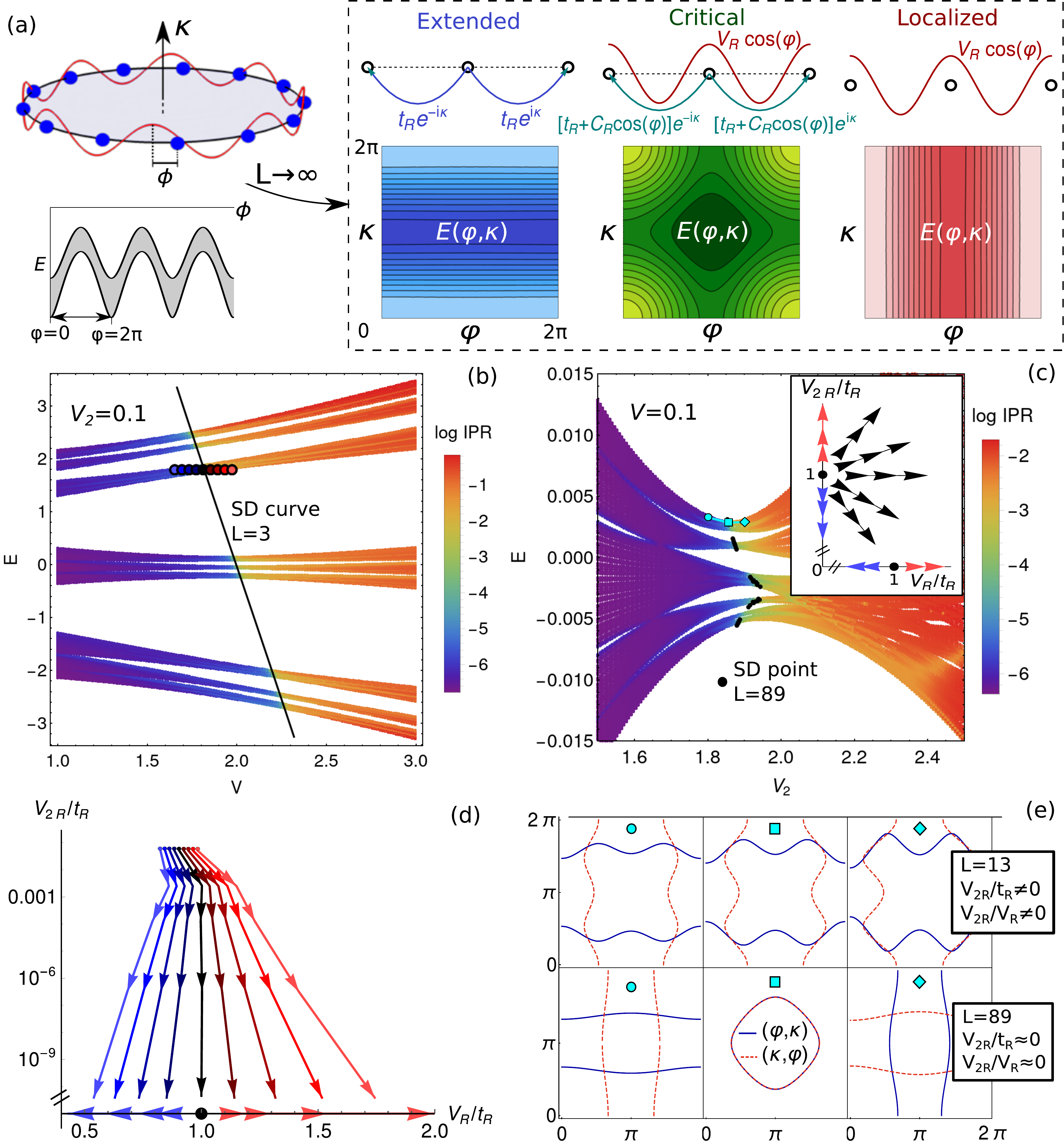}
\par\end{centering}
\caption{(a-left) CA of a quasiperiodic chain. The UC is threaded by a flux
$\kappa$, and $\phi$ is the shift between the potential (red) and
the lattice sites (blue). The periodicity of the energy with $\phi=\varphi/L$
is shown below. (a-right) Effective single-band models to which the
constant-energy contours of $E(\varphi,\kappa)$ flow as $L$ increases:
extended, with no renormalized potential; localized, with no hoppings;
and critical, with both hoppings and potential. (b-c) ${\rm IPR}$
for the model in Eq.$\,$\eqref{eq:H_V2} with $t=1$ and $\tau_{c}=610/987$.
Phase boundary computed analytically for $\tau_{c}=2/3$ in (b) {[}full
black line{]}, and numerically for $\tau_{c}=55/89$ in (c) {[}black
points{]}. Inset (c) - sketch of the RG flow close to the repulsive
fixed-point $V_{2R}/t_{R}=1$. (d) RG flow for the sequence $\tau_{c}=1,1/2,2/3,3/5,5/8$
{[}color coded by starting points in (b){]}. (e) Generalized Fermi
surfaces for $\tau_{c}=8/13$ and $\tau_{c}=55/89$, for the parameters
denoted as cyan points in~(c).\label{fig:1}}
\end{figure}

Here, we develop a renormalization group (RG) theory of one-dimensional
(1D) QPS, where the RG flow is induced by increasing unit cell (UC)
sizes of subsequent commensurate approximants (CA). Central to our
construction is the dependence of the energy bands of a CA, with UC
size $L$, on the Bloch momentum $\kappa$ (a flux thread) and on
the real-space phase shift $\phi$ {[}Fig.~\ref{fig:1}(a){]}. The
energy bands are $2\pi$-periodic in $\varphi=L\phi$ and $\kappa$
\footnote{Note that this periodicity also holds if the quasiperiodicity is in
hopping (and not on-site) terms, see \citep{HdualitiesScipost} and \citep{SM}
for detailed examples.}, allowing constant-energy contours to be expressed through an harmonic
expansion in $\varphi$ and $\kappa$ whose coefficients are renormalized
as $L$ increases, defining an RG flow. We show that around criticality
and for a given energy, different phases are characterized by RG transformations
that flows into fixed-points corresponding to renormalized single-band
models that only depend on fundamental harmonics in $\varphi$ and
$\kappa$ {[}see Fig.~\ref{fig:1}(a){]}. For these models, the potential
(hoppings) become irrelevant at an extended (localized) fixed-point,
or both the potential and hoppings are relevant at any scale, signaling
a critical fixed-point. Within our theory models are classified as:
(i) \emph{fixed-point}, when the renormalized single-band models
are the simplest possible for any CA, only containing fundamental
harmonics in $\varphi$ and $\kappa$; (ii) \emph{asymptotic}, when
the latter limit is only approached as the UC size is increased.

\paragraph*{Illustrative toy model.---}

Consider the tight-binding chain $H=-t\sum_{j}(c_{j}^{\dagger}c_{j+1}+{\rm h.c.})+\sum_{j}\mathcal{V}_{j}c_{j}^{\dagger}c_{j},$
where
\begin{multline}
\mathcal{V}_{j}=V\cos(2\pi\tau j+\phi)+V_{2}\cos[2(2\pi\tau j+\phi)]\,.\label{eq:H_V2}
\end{multline}

\noindent This is the Aubry-André model (AAM) \citep{AubryAndre}
with an additional quasiperiodic second harmonic potential of strength
$V_{2}$. For the irrational $\tau$, we take a series of CAs $\{\tau_{c}^{(n)}\},\,n=1,\cdots,\infty$
defined by $\tau_{c}^{(n)}=p/L$ ($p,L$ coprime integers), with increasing
UC size $L$. We focus on how the characteristic polynomial, $\mathcal{P}^{(n)}\left(\varphi,\kappa\right)=\det[H^{(n)}(\varphi,\kappa)-E]$,
where $H^{(n)}(\kappa,\varphi)$ is the Bloch Hamiltonian for a CA
with $\tau_{c}^{(n)}$, depends on $\kappa$ and $\varphi$. Invariance
under $\kappa\rightarrow\kappa+2\pi$ ($\varphi\rightarrow\varphi+2\pi$),
implies that $\mathcal{P}^{(n)}$ depends on $\kappa$ ($\varphi$)
only via $\cos(\kappa)$ {[}$\cos(\varphi)${]} or higher harmonics.
For the pure AAM ($V_{2}=0$), only fundamental harmonics in $\varphi$
and $\kappa$ show up,
\begin{equation}
\mathcal{P}^{(n)}(\varphi,\kappa)=t_{R}^{(n)}\cos(\kappa)+V_{R}^{(n)}\cos(\varphi)+T_{R},\label{eq:pol_AAM_V2}
\end{equation}
where $t_{R}^{(n)}=-2t^{L}$ and $V_{R}^{(n)}=(-2)^{1-L}V^{L}$ are
the \textit{renormalized couplings} for the $n$-th order CA, and
$T_{R}$ is a term independent of $\kappa$ and $\varphi$. As $L\rightarrow\infty$,
there are three fixed-points for the asymptotic behaviour of $|V_{R}^{(n)}/t_{R}^{(n)}|$:
(i) $|V_{R}^{(\infty)}/t_{R}^{(\infty)}|=0$ for $|V|<2|t|$ (extended,
renormalized potential becomes irrelevant); (ii) $|V_{R}^{(\infty)}/t_{R}^{(\infty)}|=\infty$
for $|V|>2|t|$ (localized, renormalized hopping becomes irrelevant);
and (iii) $|V_{R}^{(\infty)}/t_{R}^{(\infty)}|=1$ for $|V|=2|t|$
(critical, both renormalized potential and hopping are relevant at
any scale).

For $V_{2}\neq0$, $\mathcal{P}^{(n)}$ is given by Eq.$\,$\ref{eq:pol_AAM_V2}
with an additional second harmonic in $\varphi$, $V_{2R}^{(n)}\cos(2\varphi)$,
where $V_{2R}^{(n)}=2^{1-L}V_{2}^{L}$, and $V_{2}$ dependent $t_{R}^{(n)},V_{R}^{(n)}$
(see \citep{SM} for explicit expressions). The correlation lengths
in the extended ($\xi_{e}$) and localized ($\xi_{l}$) phases are
determined by $|V_{R}^{(n)}/t_{R}^{(n)}|=e^{-L/\xi_{e}}$ and $|t_{R}^{(n)}/V_{R}^{(n)}|=e^{-L/\xi_{l}}$,
which for $V_{2}=0$ yields well-known results. The dependence of
$V_{2R}^{(n)}$ with $L$ shows that a small $V_{2}$ introduces an
irrelevant perturbation, since the fixed point of the $L\to\infty$
flow is the same as the AAM. Remarkably, the phase boundary changes
smoothly and can be analytically approximated using a CA with small
UC, as shown in Fig.$\,$\ref{fig:1}(b). An example RG flow is shown
in Fig.$\,$\ref{fig:1}(d), color coded by starting points in Fig.$\,$\ref{fig:1}(b).

A small $V$ term added to the $V_{2}\neq0$ model, however, represents
a relevant perturbation. This is shown in the inset of Fig.$\,$\ref{fig:1}(c)
close to the $V=0$ critical point at $|V_{2R}|=|t_{R}|$. The phase
boundaries only converge when $L$ is large enough so that the model
flows to the effective single-band fixed-points. Since $L$ can be
significantly large, a complex structure of mobility edges can arise,
as shown in Fig.$\,$\ref{fig:1}(c) (black dashed line). Figure$\,$\ref{fig:1}(e)
shows a level cut of the eigen-energies, $E^{(n)}(\kappa,\varphi)$,
of the Bloch Hamiltonian of a CA. For $L\to\infty$, a localized (delocalized)
fixed point is characterized by the independence of the $E^{(n)}$
on $\kappa$ ($\varphi$), whereas $E^{(n)}(\kappa,\varphi)=E^{(n)}(\varphi,\kappa)$
at the critical point. While the contribution of $V_{2R}$ is still
clear for $L=13$, it becomes negligible for $L=89$.

This example illustrates the universal behaviour we observe in all
studied models (see also the Supplementary Material (SM) \citep{SM}):
around criticality, only the fundamental harmonics in $\varphi$ and
$\kappa$ survive as $L\to\infty$, and the nature of the perturbation
(relevant or irrelevant) determines how the phase diagram evolves
from the unperturbed case. In particular, we can conclude whether
there is one continuously connected mobility edge, as in the example
of an irrelevant perturbation shown in Fig.~\ref{fig:1}(b), or a
complex structure of mobility edges, as in the example of a relevant
perturbation shown in Fig.~\ref{fig:1}(c).

In the SM \citep{SM} we provide a more in-depth pedagogical study of the simple model in Eq.$\,$\ref{eq:H_V2}, with additional next-nearest-neighbor hoppings.

\paragraph*{General theory.---\label{sec:GeneralDescription}}

Consider a sequence of Bloch Hamiltonians $H^{(n)}(\varphi,\kappa)$
depending on $\varphi$ and $\kappa$, for each CA with $\tau_{c}^{(n)}$.
We classify models for which $\mathcal{P}^{(n)}$ at energy $E$ may
be written as \footnote{The full characteristic polynomial may factorize into products of
characteristic polynomials, each one written in the form of Eq.$\,$\ref{eq:car_pol_main}
(see \citep{SM} for examples). In this case, each sector has to be
studied independently as the energy bands associated to each polynomial
are decoupled. Nonetheless, for each sector we can always write $\mathcal{P}$
in the form of Eq.$\,$\ref{eq:car_pol_main}.}

\begin{align}
\begin{split}
&\mathcal{P}^{(n)}(E,\varphi,\kappa)\equiv \det[H^{(n)}(\varphi,\kappa)-E]\\
&= t_{R}^{(n)}(E)\cos(\kappa+\kappa_{0})+V_{R}^{(n)}(E)\cos(\varphi+\varphi_{0})\\
& +C_{R}^{(n)}(E)\cos(\kappa+\kappa'_{0})\cos(\varphi+\varphi'_{0})\\
& +\epsilon_{R}^{(n)}(E,\varphi,\kappa)+T_{R}^{(n)}(E),
\end{split}
\label{eq:car_pol_main}
\end{align}

\noindent where $\epsilon_{R}^{(n)}$ contains all the terms that
depend on $\varphi$ and $\kappa$, but are not fundamental harmonics,
and $T_{R}^{(n)}(E)$ includes all terms independent of $\varphi$
and $\kappa$. Note that $\kappa_{0},\varphi_{0},\kappa'_{0},\varphi'_{0}$
may also depend on $E$. Our central conjecture is that close enough
to phase boundaries, the term $\epsilon_{R}^{(n)}$ becomes irrelevant
with respect to the the \textit{fundamental couplings} $t_{R}^{(n)},V_{R}^{(n)}$
and $C_{R}^{(n)}$. As $n$ increases, the different phases are then
fully characterized:
\begin{equation}
\begin{aligned}\textrm{Extended: } & |C_{R}^{(n)}/t_{R}^{(n)}|,|V_{R}^{(n)}/t_{R}^{(n)}|\rightarrow0\\
\textrm{Localized: } & |C_{R}^{(n)}/V_{R}^{(n)}|,|t_{R}^{(n)}/V_{R}^{(n)}|\rightarrow0\\
\textrm{Critical: } & |C_{R}^{(n)}/t_{R}^{(n)}|,|C_{R}^{(n)}/V_{R}^{(n)}|\geq1.
\end{aligned}
\label{eq:phase_conditions}
\end{equation}
The picture in the $n\rightarrow\infty$ limit is the following:
in the extended (localized) phase only the renormalized hopping (potential),
coupled to the $\kappa$-($\varphi$-)dispersion, is relevant --
the energy dispersions flow to the effective extended (localized)
single-band model illustrated in Fig.~\ref{fig:1}(a). In the critical
phase, both the $\kappa$- and $\varphi$-dispersions are relevant,
being characterized by a flow into the critical single-band model
of Fig.~\ref{fig:1}(a).The conditions for the different phase boundaries,
including extended-to-localized (E-L), critical-to-extended (C-E),
and critical-to-localized (C-L), can be summarized as follows:
\begin{equation}
\begin{aligned}\textrm{E-L: } & |V_{R}^{(n)}/t_{R}^{(n)}|\rightarrow1\vee(|V_{R}^{(n)}/C_{R}^{(n)}|,|t_{R}^{(n)}/C_{R}^{(n)}|\rightarrow0)\\
\text{C-E: } & |C_{R}^{(n)}/t_{R}^{(n)}|\rightarrow1\wedge|C_{R}^{(n)}/V_{R}^{(n)}|\geq1\\
\text{C-L: } & |C_{R}^{(n)}/V_{R}^{(n)}|\rightarrow1\wedge|C_{R}^{(n)}/t_{R}^{(n)}|\geq1.
\end{aligned}
\label{eq:phase_boundary_conditions}
\end{equation}
Based on the nature of the $\epsilon_{R}^{(n)}$ term in Eq.$\,$\ref{eq:car_pol_main},
we can classify the 1D QPS in two different groups: \emph{fixed-point
models}, with $\epsilon_{R}^{(n)}(E,\varphi,\kappa)=0\,\forall n$,
and \emph{asymptotic models}, with $\epsilon_{R}^{(n)}\rightarrow0$
faster than the fundamental couplings $t_{R}^{(n)},V_{R}^{(n)},C_{R}^{(n)}$
(as long as they are finite), close to criticality when $n\rightarrow\infty$.Remarkably,
for fixed-point models the phase boundaries in the limit of $n\rightarrow\infty$
(quasiperiodic-limit) may be obtained through the lowest-order CA.
As shown below, the solvable models reported in Refs.~\citep{AubryAndre,PhysRevLett.104.070601,PhysRevLett.114.146601,Liu2015,Wang2020a}
are all examples of fixed-point models.

\begin{figure}[h]
\begin{centering}
\includegraphics[width=1\columnwidth]{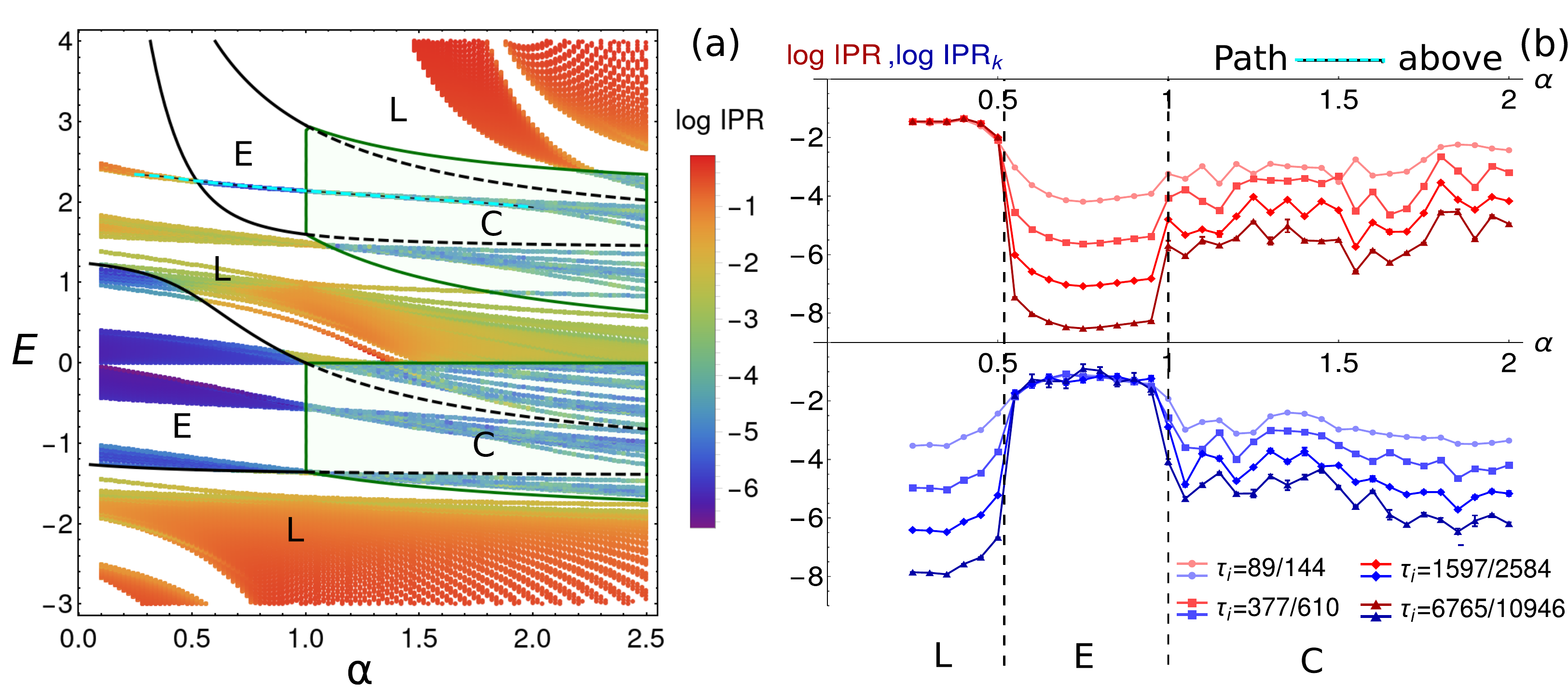}
\par\end{centering}
\caption{(a) ${\rm IPR}$ for the model in Eq.$\,$\ref{eq:crit2}, for $\tau_{c}=613/988$.
Analytical phase boundaries are superimposed as full curves: C-critical
, E-extended, and L-localized phases. SD points inside C are in dashed
black. (b) Finite-size scaling of ${\rm IPR}$ and ${\rm IPR}_{k}$
along the path shown in~(a); 10-150 averages over $\phi$ and $\kappa$
were used. \label{fig:2}}
\end{figure}

\paragraph{Fixed-point models.---}

To exemplify the application of the general description to models
beyond the AAM \citep{AubryAndre}, we consider the mosaic models
of Ref.~\citep{Wang2020a}. These are tight-binding chains with the
same potential as in the AAM, $\mathcal{V}_{j}=V\cos(2\pi\tau j+\phi)$,
but only for $j/p\in\mathbb{N}$, with $\mathcal{V}_{j}=0$ otherwise.
We consider the case $p=3$, for which the simplest possible CA has
3 sites in the UC. The $n^{{\rm th}}$-order CA has $L=3n$ sites
in the UC and $|V_{R}^{(n)}/t_{R}^{(n)}|=(|V/2||E^{2}-1|)^{n}$. The
mobility edge is simply defined by $f_{E,V}\equiv|V/2||E^{2}-1|=1$,
in complete agreement with \citep{Wang2020a}. The correlation lengths
are then $\xi_{e}=3/\log(f_{E,V}^{-1})$ and $\xi_{l}=3/\log(f_{E,V})$.

We now consider the following Hamiltonian

\begin{equation}
H=\sum_{n}\Big[t(a_{n}^{\dagger}b_{n}+b_{n}^{\dagger}a_{n+1})+{\rm h.c.}+\frac{2V\cos(2\pi\tau n+\phi)}{1+\alpha\cos(2\pi\tau n+\phi)}a_{n}^{\dagger}a_{n}\Big].\label{eq:crit2}
\end{equation}
This is a fixed-point model, which generalizes models previously studied
in Refs.~\citep{PhysRevLett.114.146601,anomScipost}. The phase
boundary between localized and delocalized regions is given by $E=\Big(V\pm\sqrt{V^{2}\pm2t^{2}\alpha+2t^{2}\alpha^{2}}\Big)/\alpha$,
while the critical phase is defined by $|\alpha(E^{2}-2t^{2})-2EV|\leq|2t\alpha|\wedge\textrm{ }|\alpha|\geq1$
\citep{SM}. The analytical phase boundaries, shown in Fig.$\,$\ref{fig:2},
are in perfect agreement with the numerical results for the ${\rm IPR}$
and ${\rm IPR}_{k}$ \footnote{We used ${\rm IPR}_{(k)}(E)=\sum_{j}|\psi_{j}^{(k)}(E)|^{4}/(\sum_{j}|\psi_{j}^{(k)}(E)|^{2})^{2}$,
where $\psi_{j}(E)$ is the wavefunction amplitude in site $j$, and
$\psi_{j}^{k}(E)$ its Fourier transform \citep{Aulbach_2004}}. In the extended (localized) phase, the ${\rm IPR}({\rm IPR}_{k})$
scales as $N^{-1}$, with $N$ the total number of sites, while in
the localized (extended) phase it is $N$-independent. At a critical
point or phase, the wavefunction is delocalized in real and momentum-space
and both the ${\rm IPR}$ and ${\rm IPR}_{k}$ scale down with $N$
\citep{Aulbach_2004}, as seen in Fig.$\,$\ref{fig:2}(b). In the SM \citep{SM} we also study an additional simpler generalization of the models in Refs.$\,$\citep{PhysRevLett.114.146601,anomScipost} and again simply unveil analyticaly the phase diagram, also including critical phases. 

In the SM \citep{SM}, we also build different classes of fixed point models.
One such class includes the models in Refs.$\,$\citep{AubryAndre,PhysRevLett.104.070601,PhysRevLett.114.146601}.
For this class, we evaluated the correlation length critical exponent
$\nu$ , obtaining $\nu_{{\rm e\text{-}l}}=1$ at the extended-localized
transition, and $\nu_{{\rm e(l)\text{-}c}}=1/2$ at extended-critical
and localized-critical transitions. We also show that our theory can be applied to even more complex cases, by studying a non-abelian Aubry-André model \citep{PhysRevB.93.104504}. We not only obtain all the phase boundaries analytically for the first time, to our knowledge, but also analytically compute the correlation length exponent for the localized-critical and extended-critical transitions to be $\nu=1$ for this model (which was recently numerically calculated for the critical-localized transition \citep{PhysRevB.106.144205}). This implies that such transitions belong to a different universality class than the analogous transitions that take place in the abelian class of models just mentioned (for which $\nu=1/2$), which can be traced back to a clear different scaling behaviour of the renormalized couplings.

\paragraph{Asymptotic models.---}

\begin{figure}[h]
\begin{centering}
\includegraphics[width=1\columnwidth]{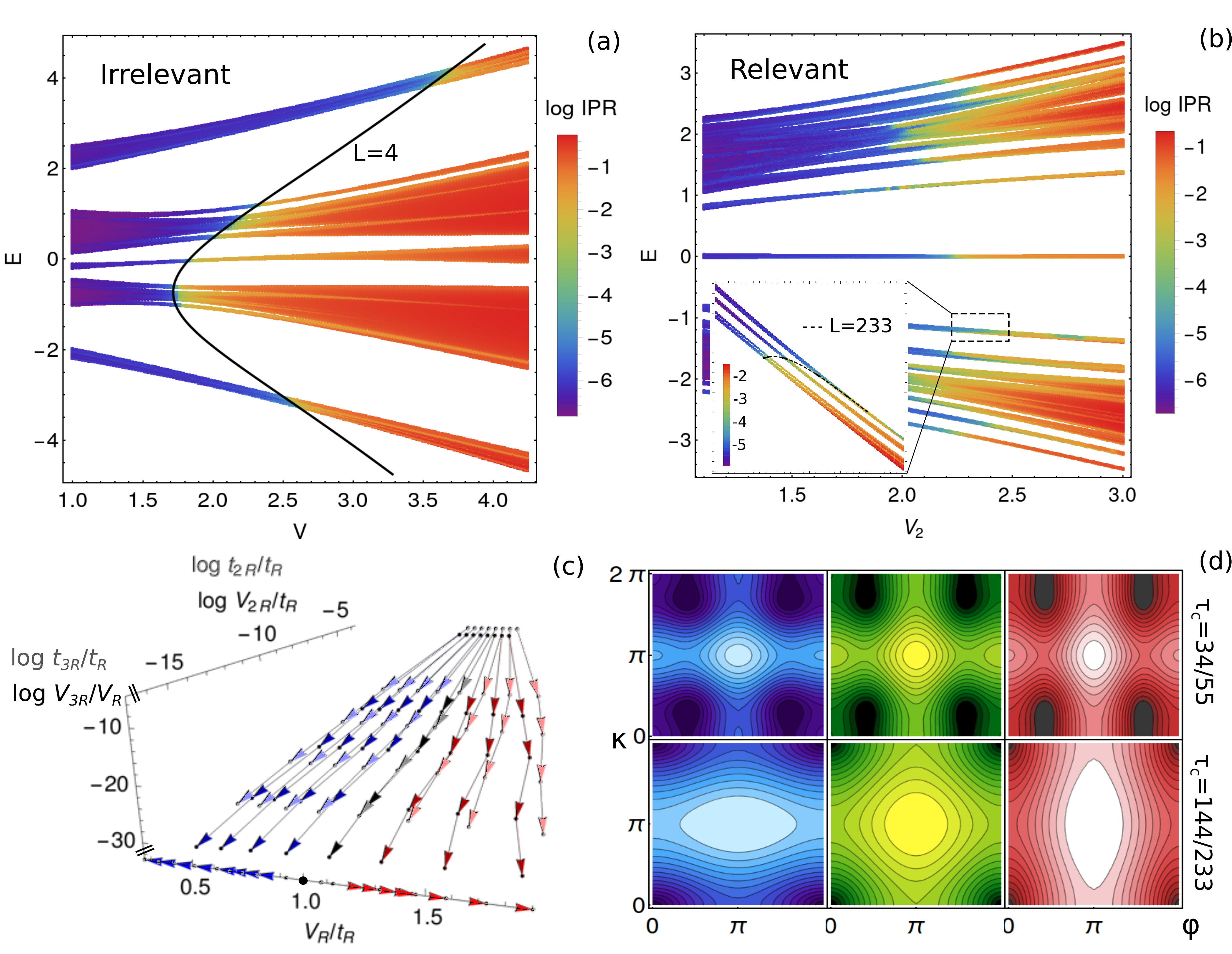}
\par\end{centering}
\caption{(a) ${\rm IPR}$ for $\tau_{c}=233/987$ along with analytical approximation
of the mobility edge for $\tau_{c}=1/4$ and parameters (irrelevant
perturbation): $t=1,t_{2}=0.1,t_{3}=0.05,V_{2}=-0.05,V_{3}=-0.075$.
(b) ${\rm IPR}$ for $\tau_{c}=233/987$ and parameters (relevant
perturbation): $t=0.05,t_{2}=1,t_{3}=V_{3}=0$. Inset: IPR for for
$\tau_{c}=987/1597$, with SD points for calculated numerically for
$\tau_{c}=144/233$. (c) Example of RG flows for $\tau_{c}=1,1/3,1/4,1/5,1/6$
and parameters (irrelevant perturbation): $t=1,t_{2}=0.025,t_{3}=0.01,V_{2}=0.025,V_{3}=0.01,E=0.5$.
(d) Contour plot of selected energy bands (close to $E=0$) for $t_{3}=V_{3}=0$
and $V=2.0$ (left), $V=2.02$ (middle), $V=2.04$ (right). Upper
panels are for CA with $L=55$ and lower panels for $L=233$ sites
in the UC. \label{fig:3}}
\end{figure}
Consider the model in Eq.$\,$\eqref{eq:H_V2}, with an additional
second-nearest-neighbor hopping term of amplitude $t_{2}$. With respect
to the AAM, the small $t_{2}$ and $V_{2}$ are irrelevant perturbations
as they do not break the original periodicity in $\varphi$ and $\kappa$
and do not change the unit cell. The higher-order harmonics of $\mathcal{P}^{(n)}$
have a very simple expression, with $\epsilon_{R}^{(n)}$ given by

\begin{equation}
\epsilon_{R}^{(n=L)}(t_{2},V_{2},\varphi,\kappa)=2t_{2}^{L}\cos(2\kappa)+2^{1-L}V_{2}^{L}\cos(2\varphi).
\end{equation}

\noindent For small $t_{2}$ and $V_{2}$, this is an example where
$\epsilon_{R}^{(n)}\rightarrow0$ exponentially faster than the fundamental
couplings as $n\rightarrow\infty$, as shown in Fig.$\,$\ref{fig:1}(d)
for $t_{2}=0$. Therefore, phase boundaries can be accurately predicted
by setting $\epsilon_{R}^{(n)}=0$\footnote{Note that one should always check whether $\epsilon_{R}$ is considerably
smaller that the renormalized couplings of the fundamental harmonics,
before making predictions for the phase diagram under that assumption.
For a given CA, it may not be accurate for all regions of the phase
diagram, if the value of the perturbation becomes significant. Away
from criticality, $\epsilon_{R}$ may even not become irrelevant as
$L\rightarrow\infty$.}. Other terms generating higher order harmonics in $\varphi$ and
$\kappa$ are also irrelevant, as illustrated in Fig.$\,$\ref{fig:3}(a)
for a model with third-nearest-neighbour hopping $t_{3}$ and potential
$V_{3}\cos(3[2\pi\tau j+\phi])$. The phase boundaries were analytically
estimated for a CA with $L=4$ sites in the UC. In Fig.$\,$\ref{fig:3}(c)
we show examples of RG flows for small perturbations and CA up to
$L=6$ sites in the unit cell.
Using the method just described, in the SM \citep{SM} we also analytically capture for the first time a reentrant localization transition unveiled in Ref.$\,$\citep{PhysRevB.105.L220201} (a similar transition was found in Ref.$\,$\citep{PhysRevLett.126.106803}).

\begin{figure}[h]
\centering{}\includegraphics[width=1\columnwidth]{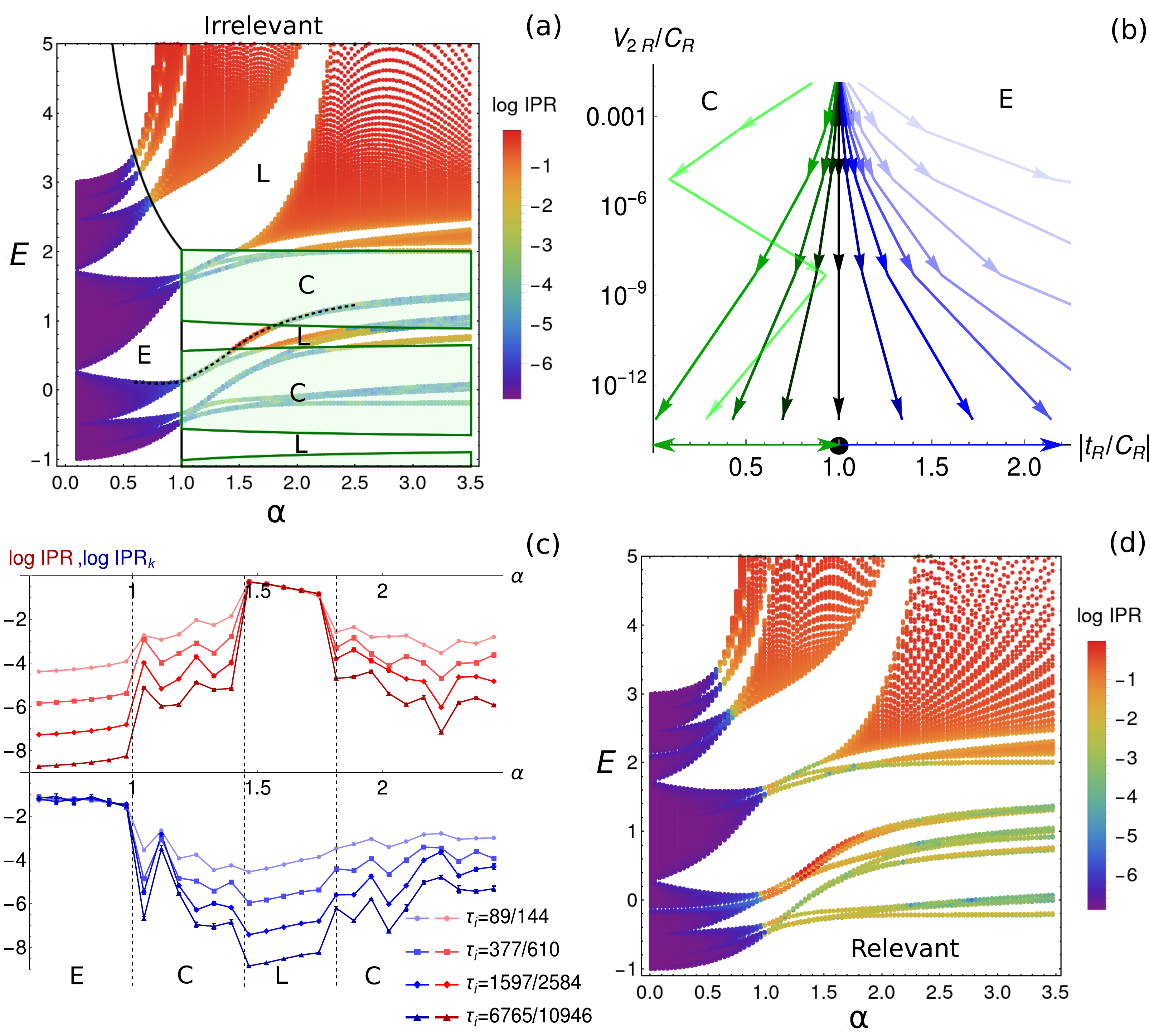}\caption{(a) ${\rm IPR}$ for the model in Eq.$\,$\eqref{eq:RP_vs_IP} with
$\tau_{c}=610/987$, $q=1$(irelevant perturbation) and $V=0.05$,
together with analytical phase boundaries for $\tau_{c}=5/8$. (b)
RG flows for $q=1,V=0.05$ in the plane $|t_{R}/C_{R}|-V_{2R}/C_{R}$,
using $\tau_{c}=1,1/2,2/3,3/5,5/8$. These results are energy independent
(only $V_{R}$, not shown, depends on energy). (c) Finite-size scaling
of ${\rm IPR}$ and ${\rm IPR}_{k}$ for the path shown as a dotted
line in (a). (d) ${\rm IPR}$ for $\tau_{c}=610/987$ and $\nu=2$
(relevant perturbation), with $V=0.05$. \label{fig:4}}
\end{figure}
 Relevant perturbations break the original periodicity in
$\varphi$ and $\kappa$ of the model. An example of a relevant perturbation
is the inclusion of finite $V$, $t$ to a model with $V_{2},t_{2}\neq0$
and $V,t=0$.\footnote{This system in only well defined for a UC with an even number of sites,
in which case it consists of two decoupled chains. The sites in a
given CA appear in pairs.}. In Fig.$\,$\ref{fig:3}(d), the generalized Fermi surface curves
for $V,t=10^{-4}$, obtained for $\tau_{c}=34/55$ (top row), qualitatively
differ from those obtained for a larger CA, $\tau_{c}=144/233$ (bottom
row), where only the fundamental harmonics are already visible \citep{SM}.
The non-trival transition lines, shown in Fig.$\,$\ref{fig:3}(b),
can still be well approximated using a large enough UC (in this case
$\tau_{c}=144/233$).

Finally, to illustrate the effects of irrelevant and relevant perturbations
on critical phases, we consider a tight-binding chain with potential
\begin{equation}
\mathcal{V}_{j}=\frac{1}{1+\alpha\cos(2\pi\tau j+\phi)}+V\cos[(2\pi\tau j+\phi)/q].\label{eq:RP_vs_IP}
\end{equation}
For $V=0$, this model reduces to the one of Refs.~\citep{PhysRevLett.114.146601,anomScipost}.
For $q=1$($q=2$), $V$ is an irrelevant (relevant) perturbation.
An excellent analytical approximation to the phase diagram, including
the critical phases, is obtained for the irrelevant perturbation,
as shown in Figs.$\,$\ref{fig:4}(a,c). In Fig.$\,$\ref{fig:4}(b)
we show RG flows in the $|t_{R}/C_{R}|-V_{2R}/C_{R}$ plane, where
the fixed-point $|t_{R}/C_{R}|=1$ separates the critical and extended
phases, with $V_{2R}$ becoming irrelevant in both. This implies that
critical phases are robust. Interestingly, the RG flow of $|t_{R}/C_{R}|$
in the critical phase oscillates between $0$ and $1$ \footnote{\noindent A similar behaviour occurs for the ratio $|V_{R}/C_{R}|$
inside the critical phase.}. This effect obtained exactly for any $L$ for $V=0$, where the
model becomes a fixed-point \citep{SM}. Here, the oscillation wavelength
can be shown to diverge at the C-E transition. This behaviour persists
at finite $V$, explaining the oscillations in the RG flows of Fig.$\,$\ref{fig:4}(b)
\footnote{In fact, the oscillating behaviour can be seen for any starting point
at the critical phase (even very close to the transition), as long
as the UC is large enough: the oscillating period diverges at the
extended-to-localized transition, where we have an invariant $|t_{R}/C_{R}|=1$,
as mentioned in the main text.}. In constrast, for the relevant perturbation with $q=2$, the critical
phase is unstable and the phase diagram becomes completely different
from the unperturbed case even for very small $V$, as suggested in
Fig.$\,$\ref{fig:4}(d). 

\paragraph{Discussion.---}

\label{sec:Discussion}

We devise a RG scheme that allows us to identify fixed-point models
of 1D QPS described by their lowest order CA and that include many
previously reported exactly solvable examples \citep{AubryAndre,Wang2020a,PhysRevLett.114.146601,anomScipost,PhysRevLett.104.070601}.

Furthermore, we identify asymptotic models whose renormalized couplings
flow to one of the fixed-point cases. Here, irrelevant perturbations
give rise to higher harmonics in phase shift $\varphi$ and twist
angle $\kappa$ whose couplings vanish at the fixed-point. Conversely,
perturbations corresponding to lower harmonics are generally relevant
and drive the RG flow to a different fixed-point. Hence, in the thermodynamic
limit, the properties of the model are determined by the stability
of the fixed-point models under different perturbations.

Analyzing the flow of asymptotic models, we show that phase boundaries
can be determined to an exponential accuracy in the size of the CA
since renormalized irrelevant couplings decay exponentially. The remarkable consequence is that for this broad class of models, the phase diagrams smoothly evolve from the exactly-solvable (fixed-point) cases and can be analytially approximated in a controlled way.

Consequently, our RG scheme explains both why (i) the phase diagram evolves
smoothly from the unperturbed case when new irrelevant couplings are added to a model; (ii) phase boundaries acquire
a complex fine-structure if the added couplings are relevant. This picture also provides insight into the stability of critical phases, which are typically fragile.

By applying our theory to different models, we unveiled multiple novel analytical results, some capturing interesting recently discovered phenomena \citep{PhysRevB.93.104504,PhysRevB.106.144205,PhysRevB.105.L220201}. Additionally, using the insights from our theory, we could even create new models with analytically solvable phase diagrams, showing peculiar features including anomalous mobility edges between critical and localized phases, with energy-dependent reentrant behaviours [Figs.$\,$\ref{fig:2}(a),\ref{fig:4}(a)].

More generically, our theory offers a qualitative and quantitative understanding of
the localization properties of widely different 1D QPS. It closes
the gap between seemingly disconnected results in the literature,
provides clear working criteria to create novel models with tailored
properties, and is a promising asset to unveil novel universality
classes of localization transitions.
\begin{acknowledgments}
MG and PR acknowledge partial support from Fundação para a Ciência
e Tecnologia (FCT-Portugal) through Grant No. UID/CTM/04540/2019.
BA, EVC and MG acknowledge partial support from FCT-Portugal through Grant
No. UIDB/04650/2020. MG acknowledges further support from FCT-Portugal
through the Grant SFRH/BD/145152/2019. BA acknowledges further support
from FCT-Portugal through Grant No. CEECIND/02936/2017. We finally
acknowledge the Tianhe-2JK cluster at the Beijing Computational Science
Research Center (CSRC), the Bob\textbar Macc supercomputer through
computational project project CPCA/A1/470243/2021 and the OBLIVION
supercomputer, through project through project HPCUE/A1/468700/2021
(based at the High Performance Computing Center - University of Évora)
funded by the ENGAGE SKA Research Infrastructure (reference POCI-01-0145-FEDER-022217
- COMPETE 2020 and the Foundation for Science and Technology, Portugal)
and by the BigData@UE project (reference ALT20-03-0246-FEDER-000033
- FEDER and the Alentejo 2020 Regional Operational Program. Computer
assistance was provided by CSRC, CENTRA/IST and the OBLIVION support
team. 

\end{acknowledgments}

\bibliographystyle{apsrev4-1}
\bibliography{1D_Hidden_SD_Paper}


\newpage

\onecolumngrid

\beginsupplement
\begin{center}
\textbf{\large{}Supplemental Material for: \vspace{0.1cm}
}{\large\par}
\par\end{center}
\large{Renormalization-Group Theory of 1D quasiperiodic lattice models with
commensurate approximants}

\vspace{0.3cm}

\tableofcontents{}

\newpage

\section{Fixed-point and asymptotic models: A simple example}

In this section, we study in more detail a simple model to illustrate
important concepts introduced in the main text, including the differences
between fixed-point and asymptotic models, and between relevant and
irrelevant perturbations. We consider the following Hamiltonian:

\begin{equation}
H=t\sum_{n}(c_{n}^{\dagger}c_{n+1}+{\rm h.c.})+t_{2}\sum_{n}(c_{n}^{\dagger}c_{n+2}+{\rm h.c.})+\sum_{n}\Big[V\cos(2\pi\tau n+\phi)+V_{2}\cos(2[2\pi\tau n+\phi])\Big]c_{n}^{\dagger}c_{n}
\end{equation}

This model was also considered in the main text, containing additional
third-nearest-neighbor hoppings and a quasiperiodic term of the type
$V_{3}\cos(3[2\pi\tau n+\phi])$. Here we drop these terms for simplicity,
but the discussion can be generalized to this and more complicated
models. The characteristic polynomial for any CA can be written as

\begin{equation}
\begin{aligned}\mathcal{P}(\varphi,\kappa)= & t_{R}^{(n)}\cos(\kappa)+V_{R}^{(n)}\cos(\varphi)+V_{2R}^{(n)}\cos(2\varphi)+t_{2R}^{(n)}\cos(2\kappa)\end{aligned}
\end{equation}

\noindent where $\varphi=L\phi$, for a CA defined by $\tau_{c}=p/L$
($p$ and $L$ co-primes). Below, we provide the renormalized coefficients
for CA up to $5$ sites in the UC:

\begin{equation}
\begin{array}{ccccccccc}
\tau_{c}=p/L &  & t_{R} &  & V_{R} &  & t_{2R} &  & V_{2R}\\
\\
1 &  & 2t &  & V &  & 2t_{2} &  & V_{2}\\
\\
1/2 &  & -2(t^{2}+2Et_{2}) &  & -\frac{1}{2^{1}}(V^{2}+4EV_{2}) &  & 2t_{2}^{2} &  & \frac{1}{2^{1}}V_{2}^{2}\\
\\
1/3;2/3 &  & 2t[t^{2}+3t_{2}(E+t_{2})] &  & \frac{1}{2^{2}}V(V^{2}+6V_{2}E+3V_{2}^{2}) &  & 2t_{2}^{3} &  & \frac{1}{2^{2}}V_{2}^{3}\\
\\
1/4;3/4 &  & -2[t^{4}+4t^{2}t_{2}(E+t_{2}) &  & \frac{1}{2^{3}}(-V^{4}-8EV^{2}V_{2}+16V_{2}^{2}t^{2} &  & 2t_{2}^{4} &  & \frac{1}{2^{3}}V_{2}^{4}\\
 &  & +t_{2}^{2}(2E^{2}-4t_{2}^{2} &  & -16V_{2}^{2}t_{2}^{2}-8E^{2}V_{2}^{2}\\
 &  & -V^{2}+V_{2}^{2})] &  & -4V^{2}V_{2}^{2}+4V_{2}^{4})\\
\\
1/5;4/5 &  & 2t\Big[t^{4}+5t^{2}(E+t_{2}) &  & \frac{1}{2^{4}}V\Big[V^{4}+5V^{2}V_{2}(2E+V_{2}) &  & 2t_{2}^{5} &  & \frac{1}{2^{4}}V_{2}^{5}\\
 &  & +5t_{2}^{2}\big(E^{2}+Et_{2}-t_{2}^{2} &  & +5V_{2}^{2}\big(4E^{2}-2(\sqrt{5}+1)t^{2}\\
 &  & -\frac{\sqrt{5}+1}{8}V^{2}+\frac{\sqrt{5}-1}{8}V_{2}^{2}\big)\Big] &  & +2(\sqrt{5}-1)t_{2}^{2}+2EV_{2}-V_{2}^{2}\big)\Big]\\
 &  & \vdots &  & \vdots &  & \vdots &  & \vdots\\
 &  &  &  &  &  & 2t_{2}^{L} &  & \frac{1}{2^{L-1}}V_{2}^{L}
\end{array}\label{eq:bichromatic_couplings}
\end{equation}

\subsection{Fixed-point models}

\begin{figure}[h]
\centering{}\includegraphics[width=0.7\columnwidth]{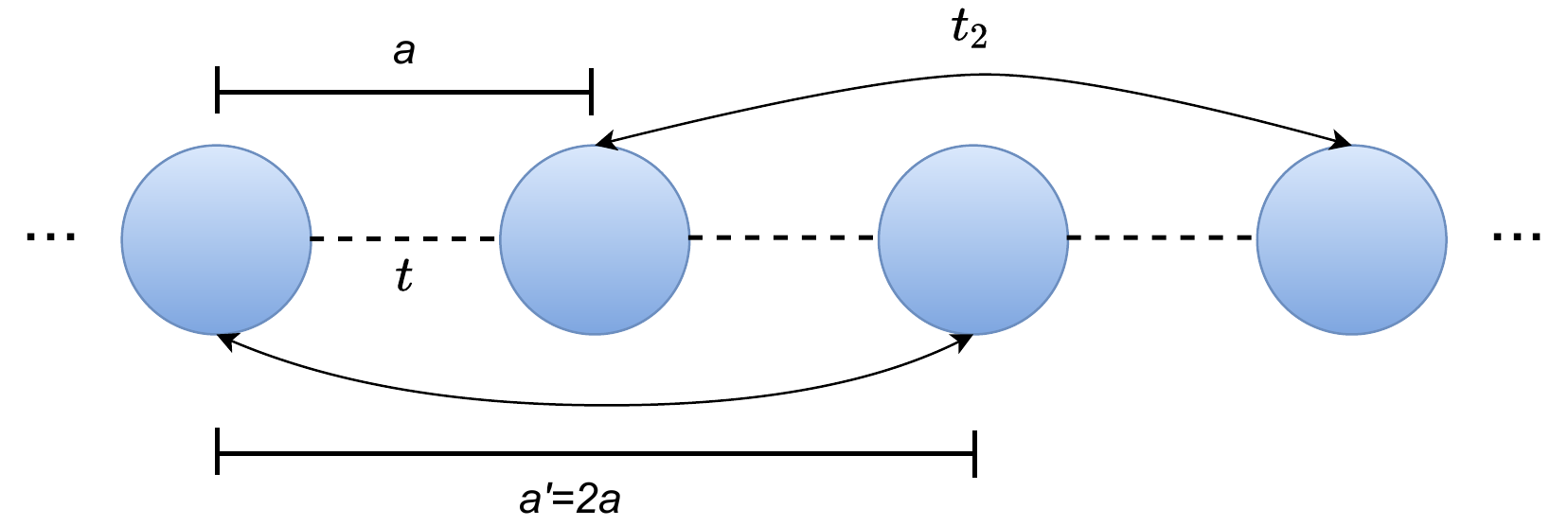}\caption{Definition of the lattice constants. $a$ is the lattice constant
for $t\protect\neq0$, while $a'$ is the lattice constant for $t=0$.
In the latter case, the system decouples into two chains.\label{fig:sketch_chains}}
\end{figure}

We first note the cases for which the model is fixed-point. In the
following discussion, we present the results in units of the lattice
constant $a$ (see Fig.$\,$\ref{fig:sketch_chains}). Note that $k$
is always the momentum measured in units of $a$.

\subsubsection{$V=t_{2}=0$ ($t-V_{2}$ model)}

In this case, if $q$ is odd, $V_{R}=0$ and only the second harmonic
survives. This is a manifestation of $V$ being a relevant perturbation
to the $t-V_{2}$ model: for these CA the unit cell is unchanged but
a lower harmonic in $\phi$ is introduced, breaking original the $\phi$-periodicity.
For even $L$, the size of the unit cell changes with respect to the
$V\neq0$ case: it contains $L/2$ sites instead of $L$. This can
be easily understood because only the potential term $V_{2}\cos(2[2\pi\tau_{c}n+\phi])$
enters the Hamiltonian. The argument of the cossine contains $2\tau_{c}=2p/L=p/(L/2)$.
If $L$ is even, the irreducible fraction occurs for $L'=L/2$, which
sets the size of the UC. This is why $V_{R}$ is finite for even $L$
in Eq.$\,$\ref{eq:bichromatic_couplings} even when $V=0$: in this
limit the calculation does not consider the true UC which gives rise
to a band flattening. As an example, let us take the $\tau_{c}=1/2$
case for $t_{2}=0$, for which the Hamiltonian matrix is

\begin{equation}
\mathcal{H}(\phi,k)=\left(\begin{array}{cc}
V\cos(\phi)+V_{2}\cos(2\phi) & 2t\cos(k)\\
2t\cos(k) & -V\cos(\phi)+V_{2}\cos(2\phi)
\end{array}\right)
\end{equation}

Of course, for $V=0$, the diagonal elements are equal, which indicates
that the unit cell has a single site in that case. It is clear that
the characteristic polynomial of the above Hamiltonian matrix will
have both $\cos(2\phi)$ and $\cos(4\phi)$ terms for $V=0$. The
correct Hamiltonian matrix for $V=0$ should be

\begin{equation}
\mathcal{H}(\phi,k)=2t\cos(k)+V_{2}\cos(2\phi)
\end{equation}

The relevance of the $V\neq0$ perturbation therefore manifests either
in the breaking of the original $\phi$ periodicity if the UC size
is unchanged (odd $L$), or in the change of the unit cell (even $L$).
The duality symmetries for $V=0$ and $V\neq0$ are therefore clearly
different.

\subsubsection{$V_{2}=t=0$ ($t_{2}-V$ model)}

For $t=0$ the system consists of two decoupled chains, that we label
$A$ and $B$ with sites separated by $a'=2a$ (see Fig.$\,$\ref{fig:sketch_chains}).
The quasiperiodic potential for each chain, in the usual units with
$a=1$, is now a term $V\cos[2\pi\tau(2n_{A(B)})+\phi]$, where $n_{A}$
and $n_{B}$ are respectively the site indexes in chains $A$ and
$B$. This case is special because the characteristic polynomial decouples
into products of polynomials corresponding to each decoupled chain
as we will see below.

The smallest-order CA of this system has one site per UC for each
chain and can be represented by the following diagonal Hamiltonian
matrix (using $\tau_{c}=1$):

\begin{equation}
\mathcal{H}_{t=0}(\phi,k)=[2t_{2}\cos(2k)+V\cos(\phi)]\mathcal{\bm{\mathcal{I}}}_{2\times2}
\end{equation}

The characteristic polynomial of course factorizes and is given by

\begin{equation}
\mathcal{P}(\phi_{1},k_{1},\phi_{2},k_{2})=\prod_{i=1}^{2}\mathcal{P}(\phi_{i},k_{i})\label{eq:pol_decomposition_t-0_V2-0}
\end{equation}

\begin{equation}
\mathcal{P}(\phi_{i},k_{i})=2t_{2}\cos(2k_{i})+V\cos(\phi_{i})-E
\end{equation}

The well-known Aubry-André critical point, $|V|=2|t_{2}|$, naturally
follows if we define $\kappa_{i}=2k_{i}$ and $\varphi_{i}=\phi_{i}$.
The characteristic polynomials for higher-order CA will of course
be products of equal polynomials, each corresponding to one Aubry-André
chain. For $t\neq0$ we get

\begin{equation}
\mathcal{H}{}_{t\neq0}(\phi,k)=\left(\begin{array}{cc}
2t_{2}\cos(2k)+V\cos(\phi) & 2t\cos(k)\\
2t\cos(k) & 2t_{2}\cos(2k)+V\cos(\phi)
\end{array}\right)
\end{equation}

However, in this case we are not specifying the correct UC. In fact,
it has only one site and the energy bands of the Hamiltonian matrix
above correspond to a band folding of the single energy band of the
correct Bloch Hamiltonian, given by

\begin{equation}
\mathcal{H}_{t\neq0}(\phi,k)=2t\cos(k)+2t_{2}\cos(2k)+V\cos(\phi)
\end{equation}

Note that the change in the UC between $t=0$ and $t\neq0$ for the
same $\tau_{c}$ is clearly a manifestation of $t$ being a relevant
perturbation.

\subsubsection{$V=t=0$ ($t_{2}-V_{2}$ model)}

This is the most dramatic case, for which both $V$ and $t$ are relevant
perturbations. The model again consists of two uncoupled chains. $t\neq0$
is clearly relevant because the UC is changed. $V\neq0$ also is because
either the UC or the periodicity in $\phi$ is changed.

The simplest CA, for $\tau_{c}=1$, corresponds to

\begin{equation}
\mathcal{H}(\phi,k)=[2t_{2}\cos(2k)+V_{2}\cos(2\phi)]\mathcal{\bm{\mathcal{I}}}_{2\times2}
\end{equation}

In this case, switching on $V$ changes the periodicity in $\phi$,
adding an additional term $V\cos(\phi)$. As in this case, we have
a term $V_{2}\cos[4\pi\tau_{c}(2n_{A(B)})+\phi]$, the $V=0$ results
are the same for $\tau_{c}=1/2$ and $\tau_{c}=1/4$. One can also
check for instance that the lowest-order CA for odd $L$ has $2L$
sites, $L$ for each chain.

Another way to see that $V$ and $t$ are relevant perturbations is
to consider CA with an odd number of sites, as show in the main text.
Such CA are only well defined when $t\neq0$. For $t=0$ every CA
should have an even number of sites: for each site in one decoupled
chain there is a pair site in the other. However, for $t=\epsilon$,
with $|\epsilon|\ll|t_{2}|$, CA with odd number of sites become well-defined.
If we choose $t=V=\epsilon$, with $|\epsilon|\ll|t_{2}|,|V_{2}|$,
we see from Eq.$\,$\ref{eq:bichromatic_couplings} that the renormalized
couplings $V_{R}$ and $t_{R}$ are very small for low-order CA. However,
as the size of the UC is increased, the fundamental harmonics should
become dominant near criticality, which implies that $V_{2R}$ and
$t_{2R}$ will become irrelevant.

\subsection{Asymptotic models: irrelevant perturbations}

For small $t_{2}$ and $V_{2}$, the couplings $V_{2R}$ and $t_{2R}$
quickly become very small with respect to $V_{R}$ and $t_{R}$, as
the UC is increased. As long as this is the case, the duality in the
fundamental harmonics of $\varphi$ and $\kappa$ may be already almost
perfect even for small UC, becoming exponentially better as the UC
is increased. For instance, for $\tau_{c}=2/3$, the analytical approximation
for the mobility edge is (solving $|V_{R}|=|t_{R}|$):

\begin{equation}
E_{c}=\frac{-8t^{3}-24tt_{2}^{2}\pm V^{3}\pm3VV_{2}^{2}}{24tt_{2}\mp6VV_{2}}
\end{equation}

\noindent which for $t_{2}=0$ corresponds to the analytical curve
plotted in Fig.$\,$1 of the manuscript. Note that there are two important
factors that may affect the accuracy of the analytical approximation.
In fact, for a more accurate approximation, we should have:
\begin{enumerate}
\item Smaller ratio between the irrelevant and relevant renormalized couplings;
\item Smaller difference between the rational approximant $\tau_{c}$ and
the irrational $\tau$, $|\tau_{c}-\tau|$.
\end{enumerate}
Note that the second factor is particulary important if the phase
boundaries have a strong dependence on $\tau$. In this case, even
if the irrelevant renormalized couplings are negligible for a given
$\tau_{c}$, the obtained phase boundaries may not be fully correct
if $\tau_{c}$ is not close enough to $\tau$. Of course, if we choose
an irrational $\tau$ very close to $\tau_{c}$, the approximation
should be very accurate regarding that point 1 is satisfied.

\begin{figure}[h]
\centering{}\includegraphics[width=1\columnwidth]{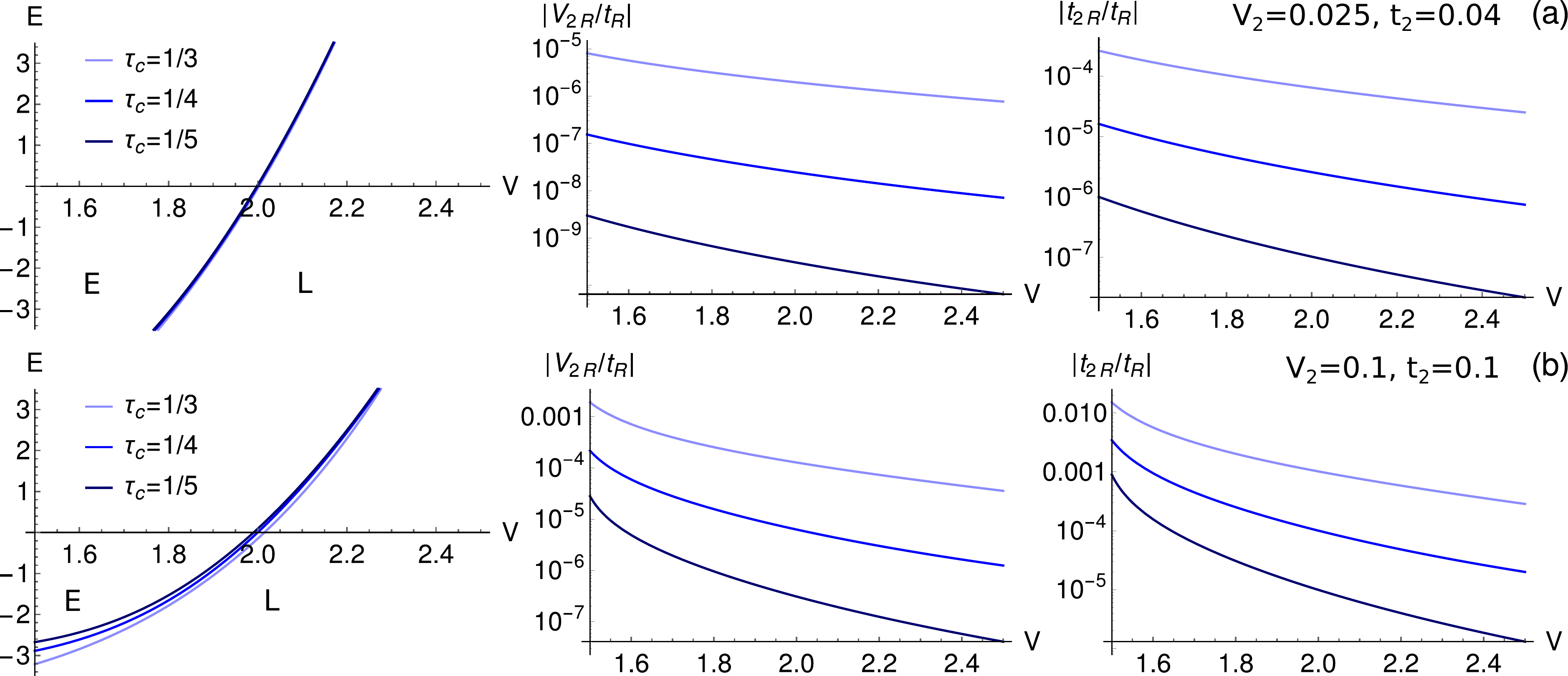}\caption{Examples of SD curves approximating transitions between an extended
(E) and localized (L) phase. In the middle and right panels we plot
the ratio between the irrelevant and relevant couplings as a function
of $V$, for energies within the SD curve (note that in this case,
$|t_{R}|=|V_{R}|$).\label{fig:sup1}}
\end{figure}

In Fig.$\,$\ref{fig:sup1} we show examples of the obtained phase
boundaries for low-order CA and the ratio between the irrelevant and
relevant couplings. Note that when $V_{2}$ and $t_{2}$ are increased,
the phase boundaries start to depend significantly on $\tau_{c}$,
even when the irrelevant couplings are negligible. This indicates
a true dependence of the phase boundaries of the limiting QPS on $\tau$.

We finish this section by emphasizing that it is always important
to make sure that the irrelevant couplings are neglegible with respect
to the relevant ones in order to make accurate analytical predictions
of the phase boundaries. In some cases, when the perturbation starts
to be significant, the renormalized terms $V_{R}$ and $t_{R}$ may
become small and higher-order approximants may be needed to describe
some phase boundaries. 

\section{Some classes of fixed-point models}

\label{sec:classes_of_fixed_point}

In this section we discuss fixed-point models for which the fundamental
renormalized couplings may be computed analytically for any CA. We
will start with the Aubry-André model and then define a class of models
that includes the latter and the models in \citep{PhysRevLett.104.070601,PhysRevLett.114.146601}.
The aim of this section is not only to show explicit examples in which
the considerations on the main text regarding fixed-point models apply,
but also to provide intuition on the important ingredients behind
the existence of these models.

\subsection{Aubry-André model}

In this section we provide additional details on fixed-point models
and how to identify them. We start with the simplest case: the Aubry-André
model. In this case, the Hamiltonian matrix for a given CA is given
by:

{\scriptsize{}
\begin{equation}
\bm{\mathcal{H}}=\left(\begin{array}{cccccc}
V\Big[\phi\Big]-E & -te^{ik} & 0 & \cdots & 0 & -te^{-ik}\\
-te^{-ik} & V\Big[2\pi\tau_{c}+\phi\Big]-E & -te^{ik} & \cdots & 0 & 0\\
0 & -te^{-ik} & V\Big[2\pi(2\tau_{c})+\phi\Big]-E & \cdots & 0 & 0\\
\vdots & \vdots & \vdots & \ddots & \vdots & \vdots\\
0 & 0 & \cdots & \cdots & V\Big[2\pi\tau_{c}(n_{1}-2)+\phi\Big]-E & -te^{ik}\\
-te^{i\kappa} & 0 & \cdots & \cdots & -te^{-ik} & V\Big[2\pi\tau_{c}(n_{1}-1)+\phi\Big]-E
\end{array}\right)\label{eq:AAM_H}
\end{equation}
}{\scriptsize\par}

\noindent where $V[x]=V\cos(x)$ and $k$ is the momentum measured
in units of the UC defined for $V=0$. Considering a CA defined by
$\tau_{c}=p/q$, with $q=L$ being the number of sites in the UC,
we have that:

\begin{equation}
\begin{aligned}\mathcal{P}(E,V,\varphi,\kappa)\equiv\det(\mathcal{\bm{\mathcal{H}}}-E\mathcal{\bm{I}})= & A_{L}\cos(\kappa)+B_{L}\cos(\varphi)+T_{R}^{(n)}(E,V)\end{aligned}
\end{equation}

\noindent with $A_{L}=-2t^{L}$, $B_{L}=(-2)^{1-L}V^{L}$, $\varphi\equiv L\phi$
and $\kappa=Lk$ ($\kappa$ is the momentum measured in units of the
true unit cell for $V\neq0$, with $L$ sites). We will now show how
this result can be easily obtained. In \citep{HdualitiesScipost} we provided
a simple physical argument on why the energy bands are periodic in
$\phi$ with period $2\pi/L$ for a given CA: shifts of $\Delta\phi=2\pi/L$
correspond just to redefinitions of the UC. Therefore $\mathcal{P}$
should be periodic with period $\Delta\varphi\equiv L\Delta\phi=2\pi$.
It is then natural that it has a term proportional to $\cos(\varphi)$.
In fact, the coefficient of this term can be computed explicitly.
The only way that a term with this periodicity can arise in the determinant
of $\mathcal{H}$ is from the product of all the diagonal terms proportional
to $V$, that is:

\begin{equation}
\prod_{n=0}^{L-1}V\cos\Big(\frac{2\pi pn}{L}+\frac{\varphi}{L}\Big)=\frac{V^{L}}{(-2)^{L-1}}[\cos(\varphi)-\cos(L\pi/2)]
\end{equation}

\noindent which readily retrieves the correct coefficient $B_{L}$.
All the terms proportional to $\cos(\varphi/r),r\in\mathbb{N}\setminus\{1\}$
should vanish by our previous argument. The only question that remains
is whether terms proportional to $\cos(r\varphi),r\in\mathbb{N}\setminus\{1\}$
can be generated. But this is not the case: that would require a product
of more than $L$ terms of the type $\cos[f(n)+\varphi/L]$. Such
a term does not exist for the Aubry-André model.

A similar argument can be used to infer the $\kappa$-dependence of
$\mathcal{P}$. In this case, we can apply the Aubry-André duality
to the eigenfunction $\bm{\psi}$, defined though $\mathcal{\bm{\mathcal{H}}}\bm{\psi}=E\bm{\psi}$,
to get

\begin{equation}
\mathcal{\bm{\mathcal{H}}}\bm{\psi}=E\bm{\psi}\leftrightarrow\mathcal{\bm{\mathcal{H}}}'\bm{f}=E\bm{f}
\end{equation}

\begin{equation}
\mathcal{\bm{\mathcal{H}}}'=\bm{U}\mathcal{\bm{\mathcal{H}}}'\bm{U}^{\dagger},\textrm{}\bm{f}=\bm{U}\bm{\psi},\textrm{}\bm{U}_{mn}=e^{i2\pi\tau mn}
\end{equation}

where

{\scriptsize{}
\begin{equation}
\bm{\mathcal{H}}'=\left(\begin{array}{cccccc}
T\Big[\kappa/L\Big]-E & \frac{V}{2}e^{i\varphi/L} & 0 & \cdots & 0 & \frac{V}{2}e^{-i\varphi/L}\\
\frac{V}{2}e^{-i\varphi/L} & T\Big[2\pi\tau_{c}+\kappa/L\Big]-E & \frac{V}{2}e^{i\varphi/L} & \cdots & 0 & 0\\
0 & \frac{V}{2}e^{-i\varphi/L} & T\Big[2\pi(2\tau_{c})+\kappa/L\Big]-E & \cdots & 0 & 0\\
\vdots & \vdots & \vdots & \ddots & \vdots & \vdots\\
0 & 0 & \cdots & \cdots & T\Big[2\pi\tau_{c}(n_{1}-2)+\kappa/L\Big]-E & \frac{V}{2}e^{i\varphi/L}\\
\frac{V}{2}e^{i\varphi/L} & 0 & \cdots & \cdots & \frac{V}{2}e^{-i\varphi/L} & T\Big[2\pi\tau_{c}(n_{1}-1)+\kappa/L\Big]-E
\end{array}\right)\label{eq:AAM_H-1}
\end{equation}
}{\scriptsize\par}

\noindent with $T[x]=-2t\cos(x)$. We only changed the basis and therefore
the characteristic polynomial should be the same. The Brillouin zone
of a given CA has a period $\Delta k\equiv\Delta\kappa/L=2\pi/L$
and therefore $\mathcal{P}$ should have this periodicity. A term
proportional to $\cos(\kappa)$ should therefore exist. The only way
to get such term is, similarly to the case of Eq.$\,$\ref{eq:AAM_H}:

\begin{equation}
\prod_{n=0}^{L-1}(-2t)\cos\Big(\frac{2\pi pn}{L}+\frac{\kappa}{L}\Big)=(-2t^{L})[\cos(\kappa)-\cos(L\pi/2)],
\end{equation}

\noindent retrieving the correct coefficient $A_{L}$. Finally, we
note that $|B_{L}/A_{L}|$ decays exponentially in the extended phase
$(|V|<2)$ and grows exponentially in the localized phase $(|V|>2)$.
We can inspect the characteristic decay lengths in each phase. Starting
in the extended phase, we have

\begin{equation}
\Bigg|\frac{B_{L}}{A_{L}}\Bigg|=\Big(\frac{|V|}{2}\Big)^{L}=e^{-\log(2/|V|)L}\equiv e^{-L/\xi_{{\rm ext}}}
\end{equation}
where $\xi_{{\rm ext}}=1/\log(2/|V|)$, defined for $|V|<2$, is precisely
the correlation length in the extended phase calculated in \citep{AubryAndre}.

In the localized phase, it is the ratio $|A_{L}/B_{L}|$ that decreases
exponentially. We have

\begin{equation}
\Bigg|\frac{A_{L}}{B_{L}}\Bigg|=\Big(\frac{2}{|V|}\Big)^{L}=e^{-\log(|V|/2)L}\equiv e^{-L/\xi_{{\rm loc}}}
\end{equation}

\noindent where $\xi_{{\rm loc}}=1/\log(|V|/2)$ is the localization
length, defined for $|V|>2$.

\subsection{A more general class of fixed-point models}

\label{subsec:general_class_of_fixed_points}

To consider more generic fixed-point models, we start with the following
generic Schrodinger equation with phase twists $k$:

\begin{equation}
h(\bm{\lambda},2\pi\tau n+\phi)u_{n}=\sum_{m}e^{-ik(n-m)}f(\bm{\lambda},|n-m|)u_{m}\label{eq:schrodinger_starting-1}
\end{equation}
where $\bm{\lambda}$ is the set of Hamiltonian parameters, we singled-out
$\tau$ as the incommensurability parameter and we define $h(\bm{\lambda},2\pi\tau n+\phi)=E-\chi(\bm{\lambda},2\pi\tau n+\phi)$
for some function $\chi$. We will take CA defined by $\tau=\tau_{c}=p/L$,
with $p$ and $L$ two co-prime integers. In this case we have that
$h_{n+rL}=h_{n};u_{n+rL}=u_{n},n=0,\cdots,L-1,{\rm }r\in\mathbb{Z}$.

We start by applying the transformation $\sum_{n=0}^{L-1}e^{i2\pi\tau_{c}n\mu}=\frac{L}{N}\sum_{n}e^{i2\pi\tau_{c}n\mu}$,
where $N$ is the total number of sites in the system. The first term
in Eq.$\,$\ref{eq:schrodinger_starting-1} becomes

\begin{equation}
\sum_{n=0}^{L-1}e^{i2\pi\tau_{c}n\mu}h(\bm{\lambda},2\pi\tau_{c}n+\phi)u_{n}=\frac{L}{N}\sum_{m}e^{i2\pi\tau_{c}m\mu}h(\bm{\lambda},2\pi\tau_{c}m+\phi)u_{m},{\rm }\label{eq:hn_basis-1}
\end{equation}

For the second term we have:

\begin{equation}
\begin{aligned}\frac{L}{N}\sum_{n}e^{i2\pi\tau_{c}n\mu}\sum_{m}e^{-ik(n-m)}f(\bm{\lambda},|n-m|)u_{m}\\
\textrm{Using \ensuremath{m'=n-m}:}\\
\frac{L}{N}\sum_{m,m'}e^{i2\pi\tau_{c}\mu m'}e^{-ikm'}f(\bm{\lambda},|m'|)e^{i2\pi\tau_{c}m\mu}u_{m}\\
\textrm{Defining \ensuremath{K(\bm{\lambda},2\pi\tau_{c}\mu-k)=\sum_{m'}e^{im'(2\pi\tau_{c}\mu-k)}f(\bm{\lambda},|m'|)}}\\
\frac{L}{N}\sum_{m}K(\bm{\lambda},2\pi\tau_{c}\mu-k)e^{i2\pi\tau_{c}m\mu}u_{m},
\end{aligned}
\label{eq:hop_basis-1}
\end{equation}

Combining, we get

\begin{equation}
\begin{aligned}\sum_{m}e^{i2\pi\tau_{c}m\mu}[E-\chi(\bm{\lambda},2\pi\tau_{c}m+\phi)-K(\bm{\lambda},2\pi\tau_{c}\mu-k)]u_{m} & =0\end{aligned}
\label{eq:M_eq-1}
\end{equation}

We now impose some restrictions in functions $\chi$ and $K$. In
particular, they should be such that Eq.$\,$\ref{eq:M_eq-1} can
be written as

\begin{equation}
\begin{aligned}\sum_{m}e^{i2\pi\tau_{c}m\mu}\Bigg(\frac{Ac_{\mu,k}+Bc_{m,\phi}+\eta c_{m,\phi}c_{\mu,k}+D}{g_{m\phi}g'_{\mu k}}\Bigg)u_{m} & =0\\
M\bm{u} & =0
\end{aligned}
\label{eq:M_eq_}
\end{equation}
where $c_{m,\phi}=\cos(2\pi\tau_{c}m+\phi)$ and $c_{\mu,k}=\cos(2\pi\tau_{c}\mu-k)$,
$\bm{u}=(u_{0},\cdots,u_{L-1})$ and $g_{m\phi}=g(2\pi\tau_{c}m+\phi)$,
$g'_{\mu k}=g'(2\pi\tau_{c}\mu-k)$ are some functions. Note that
$P_{\mu m}\equiv Ac_{\mu,k}+Bc_{m,\phi}+\eta c_{m,\phi}c_{\mu,k}+D$
contains the coefficients of the characteristic polynomial of the
1-site CA, with $\tau_{c}=1$. In particular, $P_{00}$ corresponds
exactly to this characteristic polynomial.

We will now compute the $\phi$- and $k$-dependent parts of the determinant
of matrix $M$ in Eq.$\,$\ref{eq:M_eq_}. Our final aim is to calculate
the ratios between renormalized couplings $|t_{L}/C_{L}|,|V_{L}/C_{L}|$
and $|t_{L}/V_{L}|$, where these couplings are defined, for the characteristic
polynomial of a CA with $L$ sites in the unit cell, $\mathcal{P}_{L}(\varphi,\kappa)$,
as

\begin{equation}
\begin{aligned}\mathcal{P}_{L}(\varphi,\kappa)= & V_{L}\cos(\varphi)+t_{L}\cos(\kappa)\\
 & +C_{L}\cos(\varphi)\cos(\kappa)+\cdots
\end{aligned}
\end{equation}
where $\varphi=L\phi$ and $\kappa=Lk$. The renormalized couplings
only depend on the size of the CA's unit cell for this class of models,
as we will see. We will also see why only the fundamental harmonics
in $\varphi$ and $\kappa$ appear for any CA.

The Leibniz formula for the determinant is

\begin{equation}
{\rm det}(M)=\sum_{\bm{\sigma}\in S_{L}}{\rm sgn}(\sigma)\prod_{\mu=0}^{L-1}M_{\mu,\sigma_{\mu}}
\end{equation}
where $S_{L}$ is the set of permutations of indexes $i=1,\cdots,L$.
Using Eq.$\,$\ref{eq:M_eq_}, we have

\begin{equation}
{\rm det}(M)=\frac{1}{\prod_{\mu}g_{\mu\phi}g'_{\mu k}}\sum_{\bm{\sigma}\in S_{L}}{\rm sgn}(\sigma)\prod_{\mu=0}^{L-1}e^{i2\pi\tau_{c}\mu\sigma_{\mu}}P_{\mu,\sigma_{\mu}}\propto\sum_{\bm{\sigma}\in S_{L}}{\rm sgn}(\sigma)\prod_{\mu=0}^{L-1}e^{i2\pi\tau_{c}\mu\sigma_{\mu}}P_{\mu,\sigma_{\mu}}
\end{equation}

We start by focusing on the $|t_{L}/C_{L}|$ ratio. We have

\begin{equation}
{\rm det}(M)\propto\sum_{\bm{\sigma}\in S_{L}}{\rm sgn}(\sigma)\prod_{\mu=0}^{L-1}e^{i2\pi\tau_{c}\mu\sigma_{\mu}}\Bigg(\eta c_{\mu,k}(\frac{A}{\eta}+c_{\sigma_{\mu},\phi})+Bc_{\sigma_{\mu},\phi}+D\Bigg)\label{eq:detM_2}
\end{equation}

Only products of terms $\eta c_{\mu,k}(A/\eta+c_{\sigma_{\mu},\phi})$
matter to get the terms $t_{L}\cos(\kappa)+C_{L}\cos(\kappa)\cos(\varphi)$
of $\mathcal{P}_{L}(\varphi,\kappa)$. This is because to get terms
with $L$ times the original frequency $k$ or $\phi$, we need to
have $L$ products of terms $c_{\mu,k}$ and $c_{\mu,\phi}$. The
only way to accomplish this is by multiplying $L$ terms of the type
$\eta c_{\mu,k}(A/\eta+c_{\sigma_{\mu},\phi})$. We can also ask why
terms of the type $\cos(nx)$ with $x=\phi,k$ and $n\in\mathbb{N}<L$
do not appear in $\mathcal{P}_{L}(\varphi,\kappa)$. This would give
rise to periodicity in $\phi$ and $k$, $\Delta\phi,\Delta k>2\pi/L$,
that are forbidden because the Brillouin zone boundaries are separated
by $\Delta k=2\pi/L$. In the case of phase $\phi$, shifts of $\Delta\phi=2\pi/L$
in a CA with a unit cell with $L$ sites are just re-definitions of
this unit cell \citep{HdualitiesScipost}. The energy bands are therefore
be periodic upon these shifts. Finally, we can also ask why we cannot
have terms of the type $\cos(nx)$ with $x=\phi,k$ and $n\in\mathbb{N}>L$.
The reason is that such terms would require a number of $n>L$ products
of terms $c_{\mu,k}$ and $c_{\mu,\phi}$ that do not appear in the
determinant in Eq.$\,$\ref{eq:detM_2}. Therefore, we only have fundamental
harmonics in $\varphi$ and $\kappa$, as previously stated.

To make further progress, we will need the following identity for
$p,L$ two co-prime integers \citep{Mathematica},

\begin{equation}
\prod_{\mu=0}^{L-1}\Bigg[\cos(y)-\cos\Big(\frac{2\pi p\mu}{L}+x\Big)\Bigg]=2^{1-L}\Big[\cos(Ly)-\cos(Lx)\Big],
\end{equation}
that we can apply in eq.$\,$\ref{eq:detM_2} in the case $|A/\eta|<1$
by identifying $y=A/\eta$. We get

\begin{equation}
\begin{aligned}{\rm det}(M)\propto & \gamma_{\bm{\sigma}}\Big(-\cos(L\pi/2)+\cos(Lk)\Big)\Big(\cos[L\arccos(A/\eta)]-\cos(L\phi)\Big)\\
\propto & \cos[L\arccos(A/\eta)]\cos(Lk)-\cos(Lk)\cos(L\phi)+\cdots
\end{aligned}
\end{equation}
where we identified

\begin{equation}
\gamma_{\bm{\sigma}}=\sum_{\bm{\sigma}\in S_{L}}{\rm sgn}(\sigma)\prod_{\mu=0}^{L-1}e^{i2\pi\tau_{c}\mu\sigma_{\mu}}
\end{equation}

\begin{figure}[h]
\begin{centering}
\includegraphics[width=1\columnwidth]{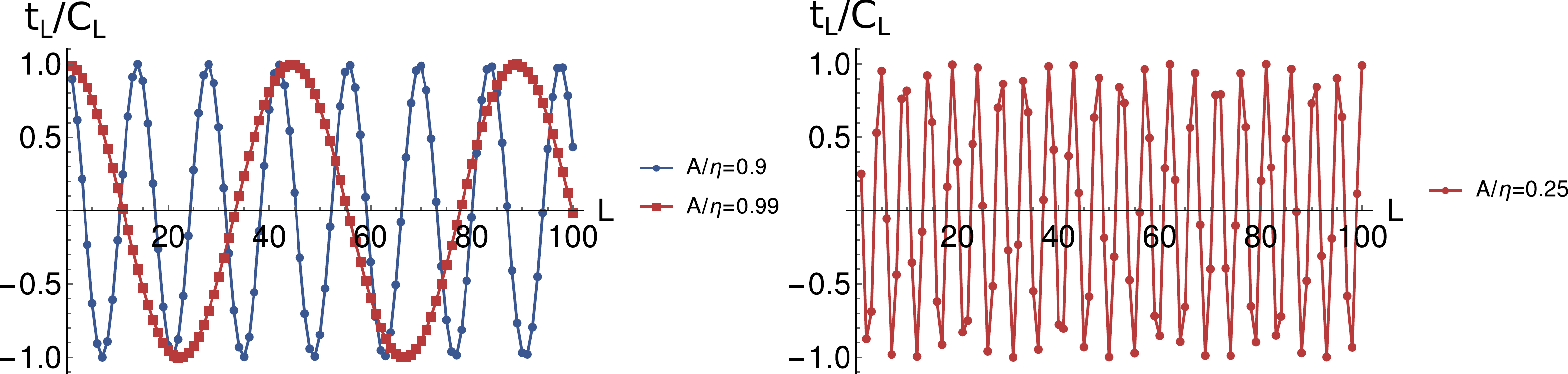}
\par\end{centering}
\caption{$t_{L}/C_{L}$ for different $A/\eta<1$.\label{fig:AR_over_etaR}}
\end{figure}

We finally have that

\begin{equation}
|t_{L}/C_{L}|=|\cos[L\arccos(A/\eta)]|=|T_{L}(A/\eta)|\label{eq:tL_CL}
\end{equation}
where $T_{L}(x)$ is the Chebyshev polynomial of order $L$. It is
very interesting to realize that there is no well-defined limit of
$|t_{L}/C_{L}|$ for large $L$. In fact, even though we always have
$|t_{L}/C_{L}|<1$, its value oscillates with $L$ with a period that
becomes larger the closer $A/\eta$ is to $1$ (diverging at $A/\eta=1$,
at the extended-to-critical transition). Examples are shown in Fig.$\,$\ref{fig:AR_over_etaR}.

We now focus in the case $|A|\geq|\eta|$. In this case, it is useful
to use the following property \citep{Mathematica}:

\begin{equation}
\prod_{\mu=0}^{L-1}\Bigg[x\pm y\cos\Big(\frac{2\pi p\mu}{L}+z\Big)\Bigg]=\frac{1}{2^{L}}\Bigg[\Bigg(x+\sqrt{x^{2}-y^{2}}\Bigg)^{L}+\Bigg(x-\sqrt{x^{2}-y^{2}}\Bigg)^{L}-2(\mp y)^{L}\cos(Lz)\Bigg]
\end{equation}

In this case we obtain 

\begin{equation}
|C_{L}/t_{L}|=\frac{2}{\Bigg(|\frac{A}{\eta}|+\sqrt{(\frac{A}{\eta})^{2}-1}\Bigg)^{L}+\Bigg(|\frac{A}{\eta}|-\sqrt{(\frac{A}{\eta})^{2}-1}\Bigg)^{L}}
\end{equation}

Note that for $|A/\eta|>1$ this gives an exponential decay for $|C_{L}/t_{L}|$.
This is just what we expect in the extended phase, where the dominant
coupling is $t_{L}$, regarding that it also dominates over $V_{L}$.
If we assume that the decay is $|C_{L}/t_{L}|\sim e^{-L/\xi_{c}}$
for large $L$, then the decay length $\xi_{EC}$ is

\begin{equation}
\begin{aligned}\xi_{EC}=\frac{1}{\log\Bigg[|A/\eta|+\sqrt{(A/\eta)^{2}-1}\Bigg]}\end{aligned}
\label{eq:corr_len_ext_crit}
\end{equation}
which is finite for any $|A/\eta|>1$ and of course diverges for $|A/\eta|=1$.
This correlation length characterizes a transition between an extended
and a critical phase. For the transition between the localized and
critical phase, we should look at the ratio $|V_{L}/C_{L}|$ that
can be computed following identical steps as before to give:

\begin{equation}
\begin{aligned}|V_{L}/C_{L}|= & \begin{cases}
|T_{L}(B/\eta)| & ,\textrm{ }|B/\eta|<1\\
\frac{1}{2}\Bigg[\Bigg(|\frac{B}{\eta}|+\sqrt{(\frac{B}{\eta})^{2}-1}\Bigg)^{L}+\Bigg(|\frac{B}{\eta}|-\sqrt{(\frac{B}{\eta})^{2}-1}\Bigg)^{L}\Bigg] & ,\textrm{ }|B/\eta|\geq1
\end{cases}\end{aligned}
\end{equation}

For $|B/\eta|\geq1$ we get the following correlation length characterizing
the localized-critical transition:

\begin{equation}
\begin{aligned}\xi_{{\rm LC}}=\frac{1}{\log\Bigg[|B/\eta|+\sqrt{(B/\eta)^{2}-1}\Bigg]}\end{aligned}
\label{eq:corr_len_loc_crit}
\end{equation}

We can finally compute $|t_{L}/V_{L}|$ through the previous expressions.
For $|V_{L}/C_{L}|,|t_{L}/C_{L}|>1$ (outside the critical phase),
we have

\begin{equation}
\begin{aligned}|t_{L}/V_{L}|= & \frac{\Bigg(|\frac{A}{\eta}|+\sqrt{(\frac{A}{\eta})^{2}-1}\Bigg)^{L}+\Bigg(|\frac{A}{\eta}|-\sqrt{(\frac{A}{\eta})^{2}-1}\Bigg)^{L}}{\Bigg(|\frac{B}{\eta}|+\sqrt{(\frac{B}{\eta})^{2}-1}\Bigg)^{L}+\Bigg(|\frac{B}{\eta}|-\sqrt{(\frac{B}{\eta})^{2}-1}\Bigg)^{L}}\end{aligned}
\label{eq:tLOverVL}
\end{equation}

Finally, similarly to what we did in the previous section for the
Aubry-André model, we can also define the correlation or localization
lengths from the scaling of the ratio $|t_{L}/V_{L}|$ with $L$,
through Eq.$\,$\ref{eq:tLOverVL}. We first assume that $|A|<|B|$
(localized phase). Taking the large $L$ limit we get

\begin{equation}
\begin{aligned}\begin{aligned}|t_{L}/V_{L}| & =e^{-L/\xi_{{\rm LE}}},\textrm{ \textrm{with }\ensuremath{\xi_{{\rm LE}}=1/}}\log\Bigg[\frac{|\frac{B}{\eta}|+\sqrt{(\frac{B}{\eta})^{2}-1}}{|\frac{A}{\eta}|+\sqrt{(\frac{A}{\eta})^{2}-1}}\Bigg]\end{aligned}
\end{aligned}
\label{eq:corr_len_loc_ext}
\end{equation}

The $\xi_{{\rm LE}}$ above is the localization length. We can also
compute the correlation length in the extended phase for $|A|>|B|$.
This can be done just by changing $A$ and $B$ in the expression
above, that is:

\begin{equation}
\begin{aligned}\begin{aligned}|V_{L}/t_{L}| & =e^{-L/\xi_{{\rm EL}}},\textrm{ \textrm{with }\ensuremath{\xi_{{\rm EL}}}=1/\ensuremath{\log\Bigg[\frac{|\frac{A}{\eta}|+\sqrt{(\frac{A}{\eta})^{2}-1}}{|\frac{B}{\eta}|+\sqrt{(\frac{B}{\eta})^{2}-1}}\Bigg]}}\end{aligned}
\end{aligned}
\label{eq:corr_len_ext_loc}
\end{equation}

In the class of models here studied, the conjectures in Eqs.$\,$4,5 of the main text
can be verified analytically.

We will finally show that the Aubry-André model and the models in
Refs.$\,$\citep{PhysRevLett.104.070601,PhysRevLett.114.146601} belong
to this class of models. For the first, we have (with phase twists
$k$):

\begin{equation}
[E+t-V\cos(2\pi\tau_{c}n+\phi)]u_{n}=t\sum_{m}e^{-ik(n-m)}e^{-p|n-m|}u_{m},
\end{equation}

We can therefore identify $\chi(V,t,2\pi\tau_{c}m+\phi)=-t+V\cos(2\pi\tau_{c}m+\phi)$
and $f(t,|m'|)=te^{-p|m'|}$. We then get

\begin{equation}
\ensuremath{K(t,2\pi\tau_{c}\mu-k)=t\sum_{m'}e^{im'(2\pi\tau_{c}\mu-k)}e^{-p|m'|}=\frac{t\sinh p}{\cosh p-\cos(2\pi\tau_{c}\mu-k)}}
\end{equation}

Inserting in Eq.$\,$\ref{eq:M_eq-1}, we have

\begin{equation}
\begin{aligned}\sum_{m}e^{i2\pi\tau_{c}m\mu}[E+t-Vc_{m,\phi}-\frac{t\sinh p}{\cosh p-c_{\mu,k}}]u_{m} & =0\\
\sum_{m}e^{i2\pi\tau_{c}m\mu}\Bigg[\frac{(E+t)\cosh p-t\sinh p+(E+t)c_{\mu,k}-V\cosh pc_{m,\phi}+Vc_{m,\phi}c_{\mu,k}}{\cosh p-c_{\mu,k}}\Bigg]u_{m} & =0
\end{aligned}
\end{equation}
which is already written in the form of Eq.$\,$\ref{eq:M_eq_}, with
$A=E+t$, $B=-V\cosh p$, $\eta=V$ and $D=(E+t)\cosh p-t\sinh p$.
The transition between the extended and localized phases occurs for
$|A|=|B|$, that is,

\begin{equation}
\cosh p=\Bigg|\frac{E+t}{V}\Bigg|\label{eq:mob_edge_biddle}
\end{equation}

For this model, critical states only appear when the condition above
is satisfied: in particular, even though $\eta\neq0$, we always have
$|B|\geq|\eta|$. Notice that the Aubry-André model can be obtained
by making the substitution $t\rightarrow2te^{p}$ and taking the large
$p$ limit. If we do so, Eq.$\,$\ref{eq:mob_edge_biddle} becomes
$|V|=2|t|$, the critical point of the Aubry-André model.

For the model in Ref.$\,$\citep{PhysRevLett.114.146601} we have
\footnote{The model in \citep{PhysRevLett.114.146601} has $\beta=1$, but we
here consider this additional parameter as well.}

\begin{equation}
\Bigg[E-2\lambda\frac{1-\beta\cos(2\pi\tau_{c}n+\phi)}{1+\alpha\cos(2\pi\tau_{c}n+\phi)}\Bigg]u_{n}=t(e^{-ik}u_{n+1}+e^{ik}u_{n-1})
\end{equation}

In this case, we have

\begin{equation}
\chi(\lambda,\alpha,2\pi\tau_{c}m+\phi)=2\lambda\frac{1-\beta\cos(2\pi\tau_{c}m+\phi)}{1+\alpha\cos(2\pi\tau_{c}m+\phi)}
\end{equation}

\begin{equation}
f(t,|m'|)=t\delta_{|m'|,1}
\end{equation}

\begin{equation}
\ensuremath{K(t,2\pi\tau_{c}\mu+k)=t\sum_{m'}e^{im'(2\pi\tau_{c}\mu-k)}f(t,|m'|)=2t\cos(2\pi\tau_{c}\mu-k)}
\end{equation}
and therefore, replacing again in Eq.$\,$\ref{eq:M_eq_}, we have

\begin{equation}
\begin{aligned}\sum_{m}e^{i2\pi\tau_{c}m\mu}[E-2\lambda\frac{1-\beta c_{m,\phi}}{1+\alpha c_{m,\phi}}-2tc_{\mu,k}]u_{m} & =0\\
\sum_{m}e^{i2\pi\tau_{c}m\mu}\Bigg[\frac{E-2\lambda+(\alpha E+2\lambda\beta)c_{m,\phi}-2tc_{\mu,k}-2t\alpha c_{m,\phi}c_{\mu,k}}{1+\alpha c_{m,\phi}}\Bigg]u_{m} & =0
\end{aligned}
\end{equation}
which is again written in the form of Eq.$\,$\ref{eq:M_eq_}, with
$A=-2t$, $B=\alpha E+2\lambda\beta$, $\eta=-2t\alpha$ and $D=E-2\lambda$.
The mobility edge condition is

\begin{equation}
2|t|=|\alpha E+2\lambda\beta|
\end{equation}

Furthermore, critical phases occur for $|\alpha|>1$ (otherwise $|\eta|\leq|A|$
always), as noticed in Ref.$\,$\citep{anomScipost}. We can also
compute the correlation lengths in Eqs.$\,$\ref{eq:corr_len_loc_crit},\ref{eq:corr_len_loc_ext},\ref{eq:corr_len_ext_loc},
obtaining the same results that were obtained in \citep{PhysRevB.105.174206}
using Avila's theory for computing Lyapunov exponents \citep{avila}.

\subsection{Alternative fixed-point models}

\label{subsec:non_abelian_AAM_fixed_point}

In order to avoid confusion, we restricted our analysis in the main
text to models with the characteristic polynomial in Eq.$\,$3.
But there can also be even more generic alternative classes of fixed-point
models, given by

\begin{equation}
\begin{aligned} & \mathcal{P}^{(n)}(E,\varphi,\kappa)=\\
 & =t_{R}^{(n)}(E)\cos[n_{1}(\kappa+\kappa_{0})]+V_{R}^{(n)}(E)\cos[n_{2}(\varphi+\varphi_{0})]\\
 & +C_{R}^{(n)}(E)\cos[n_{3}(\kappa+\kappa'_{0})]\cos[n_{3}(\varphi+\varphi'_{0})]+T_{R}^{(n)}(E)
\end{aligned}
\end{equation}
with $n_{1}\neq n_{2}\neq n_{3}$. In this section we show an instructive
example of a model of this type, for which the phase diagram can still
be exactly obtained.

We consider the non-abelian Aubry-André model in Ref.$\,$\citep{PhysRevB.93.104504},
whose Hamiltonian is given by

\begin{equation}
H=\sum_{n}(\bm{c}_{n+1}^{\dagger}T_{1}\bm{c}_{n}+{\rm h.c.})+V\sum_{n}\cos(2\pi\tau n+\phi)\bm{c}_{n}^{\dagger}T_{2}\bm{c}_{n}
\end{equation}
where $\bm{c}_{n}=[c_{n}^{(1)},c_{n}^{(2)}]^{T}$, $T_{x}=-t\sigma_{z}+it_{{\rm so}}\sigma_{y}$
and $T_{2}=V\sigma_{z}$. This model can also be recast into the Aubry-André
model with p-wave pairing \citep{PhysRevB.93.104504,PhysRevB.106.144205}.

The Schrodinger equation for this model (already with phase twists)
is

\begin{equation}
\begin{aligned}-t(e^{-ik}u_{n+1}+e^{ik}u_{n-1})+t_{{\rm so}}(e^{-ik}v_{n+1}-e^{ik}v_{n-1})+V_{n}u_{n}=Eu_{n}\\
t(e^{-ik}v_{n+1}+e^{ik}v_{n-1})-t_{{\rm so}}(e^{-ik}u_{n+1}-e^{ik}u_{n-1})-V_{n}v_{n}=Ev_{n}
\end{aligned}
\label{eq:soc_1}
\end{equation}
where $V_{n}=V\cos(2\pi\tau_{c}n+\phi)$. Summing and subtracting
the two equations, we get respectively:

\begin{equation}
\begin{aligned}-e^{-ik}(t+t_{{\rm so}})\psi_{n+1}^{-}+(-t+t_{{\rm so}})e^{ik}\psi_{n-1}^{-}+V_{n}\psi_{n}^{-}=E\psi_{n}^{+}\\
e^{-ik}(-t+t_{{\rm so}})\psi_{n+1}^{+}-(t+t_{{\rm so}})e^{ik}\psi_{n-1}^{+}+V_{n}\psi_{n}^{+}=E\psi_{n}^{-}
\end{aligned}
\label{eq:soc_1-1}
\end{equation}
where $\psi_{n}^{+}=u_{n}+v_{n}$ and $\psi_{n}^{-}=u_{n}-v_{n}$.
We now multiply by $e^{2\pi i\tau_{c}n\mu}$ and sum over $n$. Making
a change of summation variable we get $\sum_{n}e^{2\pi i\tau_{c}n\mu}\psi_{n\pm1}^{-}=\sum_{n}e^{2\pi i\tau_{c}n\mu}e^{\mp2\pi i\tau_{c}\mu}\psi_{n}^{-}$
and therefore,

\begin{equation}
\begin{aligned}\sum_{n}e^{2\pi i\tau_{c}n\mu}\Big(\Big[-e^{-i(2\pi\tau_{c}\mu+k)}(t+t_{{\rm so}})+(-t+t_{{\rm so}})e^{i(2\pi\tau_{c}\mu+k)}+V_{n}\Big]\psi_{n}^{-}-E\psi_{n}^{+}\Big)=0\\
\leftrightarrow\sum_{n}e^{2\pi i\tau_{c}n\mu}\Big(\Big[-2t\cos(2\pi\tau_{c}\mu+k)+2it_{{\rm so}}\sin(2\pi\tau_{c}\mu+k)+V_{n}\Big]\psi_{n}^{-}-E\psi_{n}^{+}\Big)=0
\end{aligned}
\label{eq:soc_1-1-1}
\end{equation}

\begin{equation}
\begin{aligned}\sum_{n}e^{2\pi i\tau_{c}n\mu}\Big(\Big[e^{-i(2\pi\tau_{c}\mu+k)}(-t+t_{{\rm so}})-(t+t_{{\rm so}})e^{i(2\pi\tau_{c}\mu+k)}+V_{n}\Big]\psi_{n}^{+}-E\psi_{n}^{-}\Big)=0\\
\leftrightarrow\sum_{n}e^{2\pi i\tau_{c}n\mu}\Big(\Big[-2t\cos(2\pi\tau_{c}\mu+k)-2it_{{\rm so}}\sin(2\pi\tau_{c}\mu+k)+V_{n}\Big]\psi_{n}^{+}-E\psi_{n}^{-}\Big)=0
\end{aligned}
\label{eq:soc_1-1-1-1}
\end{equation}

For a given CA with $L$ sites in the unit cell, we set $\psi_{n}^{\pm}=\psi_{n+L}^{\pm}$.
Moreover, $\mu=0,\cdots,L-1$, since for $\mu=L$ we return to the
$\mu=0$ equation. Our problem can then be solved by diagonalizing
a $2L\times2L$ matrix $M$, that is, $M_{\mu\alpha;n\beta}\psi_{n}^{\beta}=0$,
with $\alpha,\beta=\pm$. Defining $t_{A}=-t-t_{{\rm so}}$ and $t_{B}=-t+t_{{\rm so}}$,
we have

\begin{equation}
\bm{M}_{\mu n}=e^{2\pi i\tau_{c}n\mu}\left(\begin{array}{cc}
t_{A}e^{-i(2\pi\tau_{c}\mu+k)}+t_{B}e^{i(2\pi\tau_{c}\mu+k)}+V_{n} & -E\\
-E & t_{B}e^{-i(2\pi\tau_{c}\mu+k)}+t_{A}e^{i(2\pi\tau_{c}\mu+k)}+V_{n}
\end{array}\right)
\end{equation}

The determinant of $M$ can be explicitly written as

\begin{equation}
{\rm det}(M)=\sum_{\bm{\sigma}\in S_{L}}{\rm sgn}(\sigma)\prod_{\mu=0}^{L-1}\prod_{\alpha=\pm}M_{\mu\alpha;\sigma_{\mu\alpha}}
\end{equation}
where $S_{L}$ is the set of permutations of indexes $i=1,\cdots,2L$.
The terms that respect the $2\pi/L$ periodicity in $\phi$ and in
$k$ are formed from at least $L$ products of the $\phi$- and $k$-dependent
terms. A possible contribution arises from the products,

\begin{equation}
\begin{aligned}\Big(\sum_{\bm{\sigma}}{\rm sgn}(\sigma)\prod_{\mu}e^{2\pi i\tau_{c}\mu\sigma_{\mu}}\Big) & \Bigg[\prod_{\mu}\Big(t_{A}e^{-i(2\pi\tau_{c}\mu+k)}\Big)\prod_{n}\Big(V\cos(2\pi\tau_{c}n+\phi)\Big)+ \\
& \prod_{\mu}\Big(t_{A}e^{i(2\pi\tau_{c}\mu+k)}\Big)\prod_{n}\Big(V\cos(2\pi\tau_{c}n+\phi)\Big)\Bigg]\\
&=\gamma2^{2-L}(t_{A}V)^{L}\Big(-\cos(L\pi/2)+\cos(L\phi)\Big)\cos(Lk)
\end{aligned}
\end{equation}
where $\gamma=\sum_{\bm{\sigma}}{\rm sgn}(\sigma)\prod_{\mu}e^{2\pi i\tau_{c}\mu\sigma_{\mu}}$
and we used $\prod_{\mu}e^{i(2\pi\tau_{c}\mu+k)}=-(-1)^{L}e^{iLk}$
\citep{Mathematica}. Notice that $\prod_{\mu}F_{\mu}=\prod_{\sigma_{\mu}}F_{\sigma_{\mu}}$,
with $F_{\mu}=\Big(V\cos(2\pi\tau_{c}\mu+\phi)\Big)$ because irrespectively
of the permutation $\sigma_{\mu}$ we always get the product of all
the possible terms with $\mu=0,\cdots,L-1$. A similar contribution
arises from products of $t_{B}$-dependent and $V_{n}$ terms:

\begin{equation}
\begin{aligned}
\Big(\sum_{\bm{\sigma}}{\rm sgn}(\sigma)\prod_{\mu}e^{2\pi i\tau_{c}\mu\sigma_{\mu}}\Big)& \Bigg[\prod_{\mu}\Big(t_{B}e^{-i(2\pi\tau_{c}\mu+k)}\Big)\prod_{n}\Big(V\cos(2\pi\tau_{c}n+\phi)\Big) + 
\\ & \prod_{\mu}\Big(t_{B}e^{i(2\pi\tau_{c}\mu+k)}\Big)\prod_{n}\Big(V\cos(2\pi\tau_{c}n+\phi)\Big)\Bigg]\\
=& \gamma2^{2-L}(t_{B}V)^{L}\Big(-\cos(L\pi/2)+\cos(L\phi)\Big)\cos(Lk)
\end{aligned}
\end{equation}

For odd $L$, this gives rise to a term

\begin{equation}
2^{2-L}V^{L}[t_{A}^{L}+t_{B}^{L}]\cos(L\phi)\cos(Lk)
\end{equation}

There is also a term proportional to $\cos(2Lk)$, which can only
be obtained through the product,

\begin{equation}
\begin{aligned}\Big(\sum_{\bm{\sigma}}{\rm sgn}(\sigma)\prod_{\mu}e^{2\pi i\tau_{c}\mu\sigma_{\mu}}\Big)\prod_{\mu}\Bigg[\Big(t_{A}e^{-i(2\pi\tau_{c}\mu+k)}\Big)\Big(t_{B}e^{-i(2\pi\tau_{c}\mu+k)}\Big)+\Big(t_{A}e^{i(2\pi\tau_{c}\mu+k)}\Big)\Big(t_{B}e^{i(2\pi\tau_{c}\mu+k)}\Big)\Bigg]\\
=2\gamma(t_{A}t_{B})^{L}\cos(2Lk)
\end{aligned}
\end{equation}
Similarly, there is a term proportional to $\cos(2L\phi)$, obtained
from

\begin{equation}
\begin{aligned}\Big(\sum_{\bm{\sigma}}{\rm sgn}(\sigma)\prod_{\mu}e^{2\pi i\tau_{c}\mu\sigma_{\mu}}\Big)\prod_{\mu}\Big(V\cos(2\pi\tau_{c}\mu+\phi)\Big)\prod_{\mu}\Big(V\cos(2\pi\tau_{c}\mu+\phi)\Big)\\
=V^{2L}(-2)^{1-2L}\Big(2\cos^{2}(L\pi/2)-4\cos(L\pi/2)\cos(L\phi)+\cos(2L\phi)+1\Big)
\end{aligned}
\end{equation}

From these calculations, we can already see that for odd $L$, the
terms proportinal to $\cos(L\phi)$ and to $\cos(Lk)$ computed so
far vanish. In fact, for odd $L$, we show below that the characteristic
polynomial takes the simple form

\begin{equation}
\mathcal{P}(\varphi,\kappa)=V_{L}\cos(2\kappa)+t_{L}\cos(2\varphi)+C_{L}\cos(\kappa)\cos(\varphi)+\cdots\label{eq:pol_NNA_AAM}
\end{equation}
with $\kappa=Lk$ and $\varphi=L\phi$. Interestingly, this is a fixed-point
model only for odd $L$, different from all the other models we have
discussed. The reason why the expression above is not so straighforward
to obtain is because other products that we have not considered so
far could also give rise to additional terms proportional to $\cos(Lk)$
and $\cos(L\phi)$. We will now argue that this cannot be the case
for odd $L$. We will use the explicit $M$ matrices for $L=2$ and
$L=3$ to get some insight. These are given by

\begin{equation}
M_{L=2}=\left(\begin{array}{cccc}
f_{00} & -E & f_{01} & -E\\
-E & f'_{00} & -E & f'_{01}\\
f_{10} & -E & f_{11} & -E\\
-E & f'_{10} & -E & f'_{11}
\end{array}\right)
\end{equation}

\begin{equation}
M_{L=3}=\left(\begin{array}{cccccc}
f_{00} & -E & f_{01} & -E & f_{02} & -E\\
-E & f'_{00} & -E & f'_{01} & -E & f'_{02}\\
f_{10} & -E & f_{11} & -E & f_{12} & -E\\
-E & f'_{10} & -E & f'_{11} & -E & f'_{12}\\
f_{20} & -E & f_{21} & -E & f_{22} & -E\\
-E & f'_{20} & -E & f'_{21} & -E & f'_{22}
\end{array}\right)
\end{equation}
with $f_{\mu n}=t_{A}e^{-i(2\pi\tau_{c}\mu+k)}+t_{B}e^{i(2\pi\tau_{c}\mu+k)}+V_{n}$
and $f'_{\mu n}=t_{B}e^{-i(2\pi\tau_{c}\mu+k)}+t_{A}e^{i(2\pi\tau_{c}\mu+k)}+V_{n}$.
To produce a term proportional to $\cos(Lk)$ or $\cos(L\phi)$, we
need to have a product of at least $L$ terms $f$ or $f'$. Such
product however will have a prefactor $E^{L}$. If $L$ is odd, these
terms have to cancel out because the particle-hole symmetry of the
model ensures that the energy spectrum is symmetric around $E=0$.
We could however have the terms appearing from products of $L<n<2L$
terms $f$ or $f'$, with $n$ odd. However, in this case we would
have a prefactor $E^{2L-n}$ that also vanishes because $2L-n$ is
also odd. Finally, for $n=2L$ products, the terms $\cos(Lk)$ and
$\cos(L\phi)$ vanish as we have seen because $\cos(L\pi/2)=0$ for
odd $L$. We also note that we can only have terms $\cos(nLx),\textrm{ }n=0,1,2;\textrm{ }x=k,\phi$.
There aren't enough products of $f$ or $f'$ to generate terms with
$n>2$.

Since we have seen that for odd $L$, the characteristic polynomial
acquires the simple fixed-point form in Eq.$\,$\ref{eq:pol_NNA_AAM},
we can compute the renormalized couplings analytically. Since the
results in the thermodynamic limit do not depend on whether $L$ is
even or odd, we can therefore derive all the phase boundaries analytically
using the odd $L$ results. Combining all the results above for odd
$L$, we have

\begin{equation}
\Bigg|\frac{V_{L}}{C_{L}}\Bigg|=\frac{2^{1-2L}V^{2L}}{2^{2-L}V^{L}(t_{A}^{L}+t_{B}^{L})}=\frac{1}{2}\frac{(V/2)^{L}}{t_{A}^{L}+t_{B}^{L}}
\end{equation}

\begin{equation}
\Bigg|\frac{t_{L}}{C_{L}}\Bigg|=\frac{2(t_{A}t_{B})^{L}}{2^{2-L}V^{L}(t_{A}^{L}+t_{B}^{L})}=\frac{1}{2}\frac{(2t_{A}t_{B})^{L}}{V^{L}(t_{A}^{L}+t_{B}^{L})}
\end{equation}

Assuming $t,t_{{\rm so}}>0$, we have that that $t_{B}^{L}/t_{A}^{L}=[(t_{{\rm so}}-t)/(t_{{\rm so}}+t)]^{L}\rightarrow0$
as $L\rightarrow\infty$. In this limit we have

\begin{equation}
\Bigg|\frac{V_{L}}{C_{L}}\Bigg|\rightarrow\frac{1}{2}\Big(\Big|\frac{V}{2t_{A}}\Big|\Big)^{L}\label{eq:VL_CL_NAAM}
\end{equation}

\begin{equation}
\Bigg|\frac{t_{L}}{C_{L}}\Bigg|\rightarrow\frac{1}{2}\Big(\Big|\frac{2t_{B}}{V}\Big|\Big)^{L}\label{eq:tL_CL_NAAM}
\end{equation}

The conditions $|V_{L}/C_{L}|=1$ and $|t_{L}/C_{L}|=1$ give respectively
the conditions for the critical-to-localized and critical-to-extended
transitions. We have

\begin{equation}
\Bigg|\frac{V_{L}}{C_{L}}\Bigg|=1\leftrightarrow|V|=|2t_{A}2^{1/L}|\rightarrow2|t_{A}|\leftrightarrow|V|=2|t+t_{{\rm so}}|
\end{equation}

\begin{equation}
\Bigg|\frac{t_{L}}{C_{L}}\Bigg|1\leftrightarrow|V|=|2t_{B}\Big(\frac{1}{2}\Big)^{1/L}|\rightarrow2|t_{B}|\leftrightarrow|V|=2|t-t_{{\rm so}}|
\end{equation}

These expressions are exactly the conditions for the phase transitions
obtained numerically in Ref.$\,$\citep{PhysRevB.93.104504}.

\section{Calculating the critical exponent $\nu$ across different transitions}

In this section we show some examples of analytical calculations for
the critical exponent $\nu$, for different types of transitions,
and show how our theory can be used to unveil different universality
classes.

\subsection{Class of fixed-point models in Sec.$\,$\ref{subsec:general_class_of_fixed_points}}

We start with the class of models described in Sec.$\,$\ref{subsec:general_class_of_fixed_points}.
The results below apply to all models that fall within this class,
including the AAM and the models in Refs.$\,$\citep{PhysRevLett.104.070601,PhysRevLett.114.146601}.
For the analysis below, we will assume $A,B,\eta>0$ for simplification,
without loss of generality.

\paragraph{Extended-to-localized and localized-to-extended transitions}

We will first consider the extended-to-localized transition. The correlation
length that diverges as we approach the critical point from the extended
phase is in Eq.$\,$\ref{eq:corr_len_ext_loc}. At this phase, we
have $A>B$. Close to the transition, we have that $A=B+\epsilon$,
with $\epsilon=0^{+}$. Expanding the correlation length in Eq.$\,$\ref{eq:corr_len_ext_loc}
for small $\epsilon$, we have

\begin{equation}
\xi_{{\rm EL}}\propto\epsilon^{-1}
\end{equation}
and therefore we get $\nu=1$. A similar expansion can be made for
the localization length in Eq.$\,$\ref{eq:corr_len_loc_ext} for
$B=A+\epsilon$ with $\epsilon=0^{+}$, which means approaching the
critical point from the localized phase. In this case we again see
that

\begin{equation}
\xi_{{\rm LE}}\propto\epsilon^{-1}
\end{equation}
retrieving the same critical exponent $\nu=1$.

\paragraph{Localized-to-critical and extended-to-critical transitions}

For localized-to-critical and extended-to-critical transitions, the
divergent correlation length is in Eqs.$\,$\ref{eq:corr_len_loc_crit},\ref{eq:corr_len_ext_crit}.
Approaching the critical phase from the localized and extended phases
using respectively $B=\eta+\epsilon$ and $A=\eta+\epsilon$, and
expanding for small $\epsilon=0^{\text{+}}$, we have that

\begin{equation}
\xi_{{\rm LC}}\propto\epsilon^{-1/2}
\end{equation}

\begin{equation}
\xi_{{\rm EC}}\propto\epsilon^{-1/2}
\end{equation}

This therefore retrieves the interesting critical exponents $\nu=1/2$
for transitions into the critical phase, in agreement with the results
in \citep{PhysRevB.105.174206} for the localized-to-critical transition
using the model in \citep{PhysRevLett.114.146601}.

\paragraph{Critical-to-extended and critical-to-localized transitions}

At the critical phase, there is no correlation length that can be
defined (it is infinite by definition). However, as we have seen in
Sec.$\,$\ref{subsec:general_class_of_fixed_points} there is also
a divergent length corresponding to the period of oscillations of
$|t_{L}/C_{L}|=|T_{L}(A/\eta)|$ and $|V_{L}/C_{L}|=|T_{L}(B/\eta)|$,
respectively at the critical-to-extended and critical-to-localized
transitions. If we expand $A=B=\eta-\epsilon$ for small $\epsilon$,
we can easily see that the wavelength of oscillation diverges with
$\epsilon^{-1/2}$, which has curiously the same exponent as the correlation
length exponent $\nu$ for the transitions into the critical phase.

\subsection{Model in Sec.$\,$\ref{subsec:non_abelian_AAM_fixed_point}}

From the ratios between the renormalized couplings in Eqs.$\,$\ref{eq:VL_CL_NAAM},\ref{eq:tL_CL_NAAM},
we can easily extract the critical exponent $\nu$ for the localized-to-critical
and extended-to-critical transitions. For the former, we have:

\begin{equation}
\begin{aligned}\Bigg|\frac{C_{L}}{V_{L}}\Bigg|=2\Big(\Big|\frac{2t_{A}}{V}\Big|\Big)^{L} & =2e^{-L\log(|\frac{V}{2t_{A}}|)}\end{aligned}
\end{equation}
and therefore $\xi=1/\log(|\frac{V}{2t_{A}}|)$. For $V=2t_{A}+\epsilon$,
with $\epsilon\rightarrow0^{+}$ we have $\xi\propto\epsilon^{-1}$
and therefore $\nu=1$. This exponent was only obtained numerically
very recently \citep{PhysRevB.106.144205}. Simple calculations show
that we have $\nu=1$ also for the extended-to-critical transition,
a result that was not yet obtained, to our knowledge.

\section{Similarities between three models in the literature}

We here discuss interesting similarities between the models in \citep{PhysRevLett.104.070601,PhysRevLett.114.146601,Liu2015}.
The Schrodinger equation for the model in \citep{Liu2015} can be
written as

\begin{equation}
V_{1}\cos\big[2\pi\tau_{c}n+\phi\big]\psi_{n}+\big(t+V_{2}\cos\big[2\pi\tau_{c}(n+1/2)+\phi\big]\big)\psi_{n+1}+\big(t+V_{2}\cos\big[2\pi\tau_{c}(n-1/2)+\phi\big]\big)\psi_{n-1}=E\psi_{n}\label{eq:Thouless_Model}
\end{equation}

\noindent with $\tau_{c}=p/L$, $n=0,\cdots,L-1$ and $\psi_{n+L}=\psi_{n}e^{i\kappa}$.
Below we show the couplings $V_{R}^{(n)},t_{R}^{(n)}$ and $C_{R}^{(n)}$
for different CA with different UC sizes ($L$) \footnote{Note that here we are only showing the results for $\tau_{c}=1/q$.
More generically, for $\tau_{c}=p/q$, the results remain unchanged,
except for $C_{R}^{(n)}$ for which the sign may change. For instance,
for $\tau_{c}=2/3$, $C_{R}=\frac{2}{2^{2}}V_{2}^{3}$, while for
$\tau_{c}=1/3$ it is symmetric. Nonetheless, the absolute values
(the only values that matter to obtain the phase boundaries) remain
uchanged.}:

\begin{equation}
\begin{array}{ccccccc}
\tau_{c}=1/L &  & t_{R}^{(n\equiv L)} &  & V_{R}^{(n\equiv L)} &  & C_{R}^{(n\equiv L)}\\
\\
1 &  & 2t &  & V_{1} &  & -2V_{2}\\
\\
1/2 &  & \frac{2}{2^{1}}\big[-2t^{2}+V_{2}^{2}\big] &  & \frac{1}{2}(-V_{1}^{2}+2V_{2}^{2}) &  & -\frac{2}{2^{1}}V_{2}^{2}\\
\\
1/3 &  & \frac{2}{2^{2}}\big[4t^{3}-3tV_{2}^{2}\big] &  & \frac{1}{2^{2}}(V_{1}^{3}-3V_{1}V_{2}^{2}) &  & -\frac{2}{2^{2}}V_{2}^{3}\\
\\
1/4 &  & \frac{2}{2^{3}}\big[-8t^{4}+8t^{2}V_{2}^{2}-V_{2}^{4}\big] &  & \frac{1}{2^{3}}(-V_{1}^{4}+4V_{1}^{2}V_{2}^{2}-2V_{2}^{4}) &  & -\frac{2}{2^{3}}V_{2}^{4}\\
\\
1/5 &  & \frac{2}{2^{4}}\big[16t^{5}-20t^{3}V_{2}^{2}+5tV_{2}^{4}\big] &  & \frac{1}{2^{4}}(V_{1}^{5}-5V_{1}^{3}V_{2}^{2}+5V_{1}V_{2}^{4}) &  & -\frac{2}{2^{4}}V_{2}^{5}\\
\\
1/6 &  & \frac{2}{2^{5}}\big[-32t^{6}+48t^{4}V_{2}^{2}-18t^{2}V_{2}^{4}+V_{2}^{6}\big] &  & \frac{1}{2^{5}}(-V_{1}^{6}+6V_{1}^{4}V_{2}^{2}-9V_{1}^{2}V_{2}^{4}+2V_{2}^{6}) &  & -\frac{2}{2^{5}}V_{2}^{6}
\end{array}
\end{equation}

This is a fixed-point model, as only the couplings shown above are
generated for each CA. Having realized this, we can apply the conditions
of the main text to get the phase boundaries to find the localization-delocalization
transition at $|V_{1}|=2|t|$ from $|V_{R}^{(n)}/t_{R}^{(n)}|=1$
and the critical phase at $(|V_{2}|>|V_{1}|/2)\wedge(|V_{2}|>1)$
from $|C_{R}^{(n)}/t_{R}^{(n)}|\geq1\wedge|C_{R}^{(n)}/V_{R}^{(n)}|\geq1$.

In fact, inspection of the coefficients for any CA suggests that we
can write $t_{R}^{(n)}=\mathcal{Q}_{n}^{1}(V_{2},-2t)$, $V_{R}^{(n)}=\mathcal{Q}_{n}^{1}(V_{2},-V_{1})$
and $|C_{R}^{(n)}|=|\mathcal{Q}_{n}^{2}(V_{2})|$, with

\begin{equation}
\begin{aligned}\mathcal{Q}_{L}(x,y,\varphi) & =-2\sum_{\mu=0}^{L}(y/2)^{L-\mu}\mathcal{P}_{S}^{\mu}(\{v_{1},\cdots,v_{L-1}\})\\
 & \equiv\mathcal{Q}_{L}^{1}(x,y)+\mathcal{Q}_{L}^{2}(x)\cos(\varphi)
\end{aligned}
\label{eq:pol_symmetric-1}
\end{equation}

\begin{equation}
v_{\mu}=x\cos(2\pi\tau_{c}\mu+\varphi/L),\mu=0,\cdots,L-1
\end{equation}

\noindent where $\mathcal{P}_{S}^{\mu}(\{v_{1},\cdots,v_{L-1}\})$
is the $\mu$-th order symmetric polynomial in the variables $\{v_{1},\cdots,v_{L-1}\}$.
$|\mathcal{Q}_{n}^{2}(x)|=2^{1-L}x^{L}$, while $\mathcal{Q}_{n}^{1}$
is a more complicated polynomial, as seen above for different CA.
$|\mathcal{Q}_{n}^{1}(V_{2},2)/\mathcal{Q}_{n}^{1}(V_{2},V_{1})|=1$
implies that $|V_{1}|=2$ for any CA. The critical phases can be found
through $|\mathcal{Q}_{n}^{2}(V_{2})/\mathcal{Q}_{n}^{1}(V_{2},V_{1})|>1\wedge|\mathcal{Q}_{n}^{2}(V_{2})/\mathcal{Q}_{n}^{1}(V_{2},2)|>1$,
yielding the correct result.

Remarkably, the polynomial $\mathcal{Q}_{L}$ appears in more models
in the literature whose phase diagrams host exact solutions. These
include the models in Refs.$\,$\citep{PhysRevLett.104.070601,PhysRevLett.114.146601}.
For the model in \citep{PhysRevLett.114.146601}, we have

\begin{equation}
\begin{aligned}2\lambda\frac{1-\cos(2\pi\tau_{c}n+\phi)}{1+\alpha\cos(2\pi\tau_{c}n+\phi)}\psi_{n}+t(\psi_{n+1}+\psi_{n-1}) & =E\psi_{n}\\
\leftrightarrow\frac{2\lambda-(2\lambda+\alpha E)\cos(2\pi\tau_{c}n+\phi)]}{1+\alpha\cos(2\pi\tau_{c}n+\phi)}\psi_{n}+t[\psi_{n+1} & +\psi_{n-1}]\frac{1+\alpha\cos(2\pi\tau_{c}n+\phi)}{1+\alpha\cos(2\pi\tau_{c}n+\phi)}=\frac{E}{1+\alpha\cos(2\pi\tau_{c}n+\phi)}\psi_{n}
\end{aligned}
\label{eq:dasSarma_model}
\end{equation}

Note that the second equation gives rise to a non-hermitian matrix.
Nonetheless, its solutions are the same as for the first equation
and similarities with Eq.$\,$\ref{eq:Thouless_Model} become more
apparent. This is the case if we throw away the common denominator,
which can be done as long as $1+\alpha\cos(2\pi\tau_{c}n+\phi)\ne0$.
In this case, the coefficients of $\mathcal{P}$ are the same as the
ones above up to overall signs if $V_{1}\rightarrow-(2\lambda+\alpha E)$
and $V_{2}\rightarrow\alpha t$. We therefore have that $t_{R}^{(n)}=\mathcal{Q}_{n}^{1}(\alpha t,-2t)$,
$V_{R}^{(n)}=\mathcal{Q}_{n}^{1}(\alpha t,2\lambda+\alpha E)$ and
$C_{R}^{(n)}=|\mathcal{Q}_{n}^{2}(\alpha t)|$ (as long as $1+\alpha\cos(2\pi\tau_{c}n+\phi)\neq0$).
Note that in this case, $|C_{R}^{(n)}/t_{R}^{(n)}|>1$ implies that
$|\alpha|>1$. This is why no critical phases were observed in \citep{PhysRevLett.114.146601}
because there only $|\alpha|<1$ was considered.

Finally, for the model in \citep{PhysRevLett.104.070601}, we have

\begin{equation}
\begin{aligned}t'\sum_{l}e^{-pl}\Big(e^{ilk}f_{m-l}+e^{-ilk}f_{m+l}\Big)+V\cos(2\pi\tau mp+\phi)f_{m}=Ef_{m}\end{aligned}
\label{eq:ExpDecHopsModel}
\end{equation}

Note that we are considering the phase twists ($k\equiv\kappa/L$)
uniformly distributed in the hoppings for reasons that will become
apparent. To check the similarity with the model in Eq.$\,$\ref{eq:dasSarma_model},
we notice that it can also be written as \citep{PhysRevLett.114.146601}
(again uniformly distributing the phase twists in the hoppings):

\begin{equation}
t(\psi_{n+1}e^{ik}+\psi_{n-1}e^{-ik})+2g(\lambda,\beta)\sum_{l}e^{-\beta l}\cos[l(2\pi\tau n+\phi)]\psi_{n}-4\lambda\cosh(\beta)\psi_{n}=E\psi_{n}
\end{equation}

\noindent where $g=2\lambda(1+\cosh\beta)/\sinh\beta$ and $\cosh\beta=\alpha^{-1}$.
Applying the Aubry-André duality,

\begin{equation}
\psi_{n}=\sum_{m}e^{i2\pi\tau mn}f_{m}
\end{equation}

\noindent we arrive at

\begin{equation}
\begin{aligned}g(\lambda,\beta)\sum_{l}e^{-\beta l}\Big(e^{i\phi l}f_{m-l}+e^{-i\phi l}f_{m+l}\Big)+2t\cos(2\pi\tau mp+k)f_{m}=[E+4\lambda\cosh(\beta)]f_{m}\end{aligned}
\label{eq:dasSarmaDual}
\end{equation}

We can now see that Eqs.$\,$\ref{eq:ExpDecHopsModel},\ref{eq:dasSarmaDual}
are the same if in the first we make $t'\rightarrow g(\lambda,\beta)$,
$V\rightarrow2t$, $E\rightarrow E+4\lambda\cosh(\beta)$ and $k\leftrightarrow\phi$.
Note that this last replacement is fundamental because it means that
the $V_{R}^{(n)}$ coefficient in the model of Eq.$\,$\ref{eq:ExpDecHopsModel}
is the $t_{R}^{(n)}$ coefficient in the model of Eq.$\,$\ref{eq:dasSarmaDual}
and vice-versa.

\section{More on fixed-point models with critical phases}

In this section we provide additional details on the fixed-point model
with a critical phase considered in the main text and discuss a related
model already considered in the literature.

For the fixed-point model in Eq.$\,$6 of the main text,
the lowest-order approximant has $2$ sites in the UC. The characteristic
polynomial in that case is, defining $\varphi=L\phi/2$:

\begin{equation}
\mathcal{P}(\varphi,\kappa)=\frac{E^{2}-2t^{2}-(2t^{2}\alpha+2EV-E^{2}\alpha)\cos(\varphi)+2t^{2}\cos(\kappa)+2t^{2}\alpha\cos(\kappa)\cos(\varphi)}{1+\alpha\cos(\varphi)}
\end{equation}

Note that as long as $1+\alpha\cos(\varphi)\neq0$, the energy bands
are obtained by setting the numerator to zero. The denominator diverges
at certain points $\varphi^{*}$ when $|\alpha|>1$. But for any other
$\varphi$ we may identify $\mathcal{P}$ with the numerator solely
and apply our description. In that case, applying the conditions for
the phase boundaries in Eq.$\,$5
of the main text immediately retrieves the analytical expression for
the mobility edges given below Eq.$\,$6. Remarkably,
for this model, similarly to previously discussed fixed-point models,
the phase boundaries obtained in are independent of the CA.

We now also consider the model introduced in \citep{PhysRevLett.114.146601},
whose critical phases were studied in Ref.$\,$\citep{anomScipost}.
The Hamiltonian is

\begin{equation}
H=-t\sum_{n}(c_{n}^{\dagger}c_{n+1}+c_{n+1}^{\dagger}c_{n})+\sum_{n}\frac{V_{1}+V_{2}\cos(2\pi\tau n+\phi)}{1+\alpha\cos(2\pi\tau n+\phi)}c_{n}^{\dagger}c_{n}\label{eq:crit1-1-1}
\end{equation}

For this model, the phase diagram can again be obtained by studying
the lowest-order CA ($\tau_{c}=1$), whose characteristic polynomial
is given by

\begin{equation}
\mathcal{P}(\varphi,\kappa)=\frac{V_{1}-E+(V_{2}-E\alpha)\cos(\varphi)+2t\cos(\kappa)+2t\alpha\cos(\kappa)\cos(\varphi)}{1+\alpha\cos(\varphi)}
\end{equation}

This characteristic polynomial gives rise to the following phase boundaries:

\begin{equation}
\begin{cases}
E=(V_{2}\pm2t)/\alpha & \textrm{, SD points}\\
|V_{2}-E\alpha|\leq|2t\alpha|\wedge|\alpha|\geq1 & \textrm{, Critical phase}
\end{cases}\label{eq:H1-1}
\end{equation}

\begin{figure}[h]
\begin{centering}
\includegraphics[width=0.75\columnwidth]{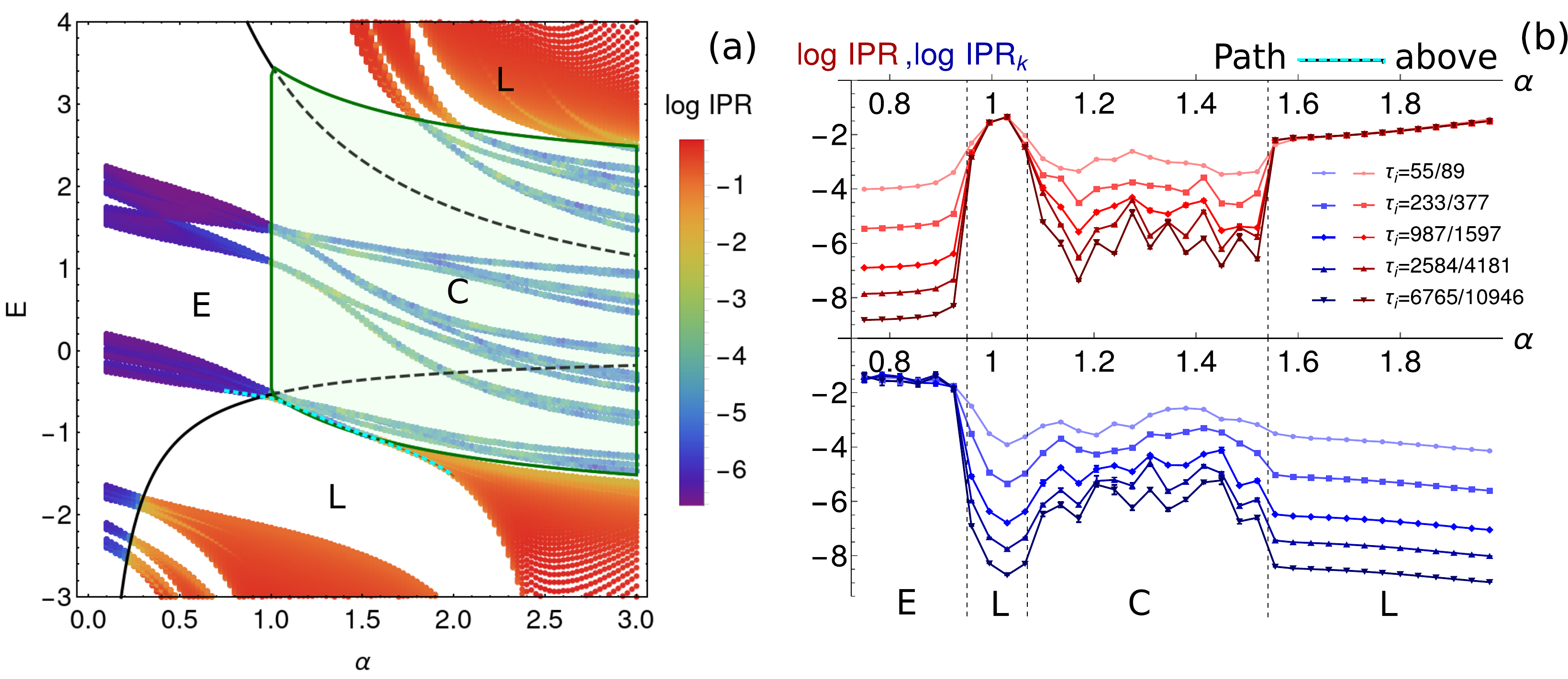}
\par\end{centering}
\caption{(a) ${\rm IPR}$ for the model in Eq.$\,$\ref{eq:crit1-1-1}(a),
for $\tau_{i}=610/987$. The analytical predictions for the phase
boundaries are superimposed: critical phase (C) in green, SD curves
at extended(E)-localized(L) transitions in full black and SD points
inside C in dashed black. (b) Finite-size scaling of ${\rm IPR}$
and ${\rm IPR}_{k}$ for paths in the phase diagram shown in the dotted
cyan lines of (a). The eigenstate with energy closer to this chosen
path was considered for each point and averages over $\phi$ and phase
twists $k$ were made (using from 10 to 150 configurations, respectively
from the larger to the smallest depicted system size).\label{fig:2-1}}
\end{figure}

We can check that for other approximants the same phase boundaries
can be obtained. For instance, for $\tau_{c}=1/2$ we have, using
$\varphi=2\phi$:

\begin{equation}
\mathcal{P}(\varphi,\kappa)=\frac{(V_{2}^{2}-2t^{2}\alpha^{2}-2EV_{2}\alpha+E^{2}\alpha^{2})\cos(\varphi)+4t^{2}\cos(\kappa)-2t^{2}\alpha^{2}\cos(\kappa)\cos(\varphi)}{-2+\alpha^{2}+\alpha^{2}\cos(\varphi)}
\end{equation}

\noindent again giving rise to the phase boundaries in Eqs.$\,$\ref{eq:H1-1}.
In Fig.$\,$\ref{fig:2-1} we exemplify how the results for the phase
boundaries obtained here perfectly match the numerical results of
the IPR.

\section{Dual points: generalization to include critical phases}

In \citep{HdualitiesScipost} we have shown how to define dual points close
to criticality for a CA of large enough order in extended and localized
phases (and at the self-dual critical point separating them). Here
we generalize our definition to include critical phases, by considering
also the renormalized coupling $C_{R}^{(n)}$.

In the main text we have written the characteristic polynomial for
a given $n^{{\rm th}}$-order CA as

\begin{equation}
\begin{aligned}\mathcal{P}(E,\bm{\lambda},\varphi,\kappa)= & t_{R}^{(n)}(E,\bm{\lambda})\cos(\kappa+\kappa_{0})+V_{R}^{(n)}(E,\bm{\lambda})\cos(\varphi+\varphi_{0})\\
 & +C_{R}^{(n)}(E,\bm{\lambda})\cos(\kappa+\kappa_{0})\cos(\varphi+\varphi_{0})\\
 & +T_{R}^{(n)}(E,\bm{\lambda})+\epsilon_{R}^{(n)}(E,\bm{\lambda},\varphi,\kappa)
\end{aligned}
\end{equation}

In the regime that $\epsilon_{R}^{(n)}\approx0$, the renormalized
couplings $t_{R}^{(n)},V_{R}^{(n)}$ and $C_{R}^{(n)}$ enable us
to find dual points. After setting $\mathcal{P}$ to zero, we get
(dropping the index $n$ for clarity):

\begin{equation}
\cos(\varphi+\varphi_{0})=\frac{T_{R}-t_{R}\cos(\kappa+\kappa_{0})}{V_{R}+C_{R}\cos(\kappa+\kappa_{0})}=\frac{\frac{T_{R}}{V_{R}}-\frac{t_{R}}{V_{R}}\cos(\kappa+\kappa_{0})}{1+\frac{C_{R}}{V_{R}}\cos(\kappa+\kappa_{0})}
\end{equation}

\noindent or

\begin{equation}
\cos(\kappa+\kappa_{0})=\frac{T_{R}-V_{R}\cos(\varphi+\varphi_{0})}{t_{R}+C_{R}\cos(\varphi+\varphi_{0})}=\frac{\frac{T_{R}}{t_{R}}-\frac{V_{R}}{t_{R}}\cos(\varphi+\varphi_{0})}{1+\frac{C_{R}}{t_{R}}\cos(\varphi+\varphi_{0})}
\end{equation}

\noindent Dual points $(E,\bm{\lambda})$ and $(E',\bm{\lambda}')$
should then satisfy:

\begin{equation}
\begin{cases}
\frac{t_{R}(E,\bm{\lambda})}{V_{R}(E,\bm{\lambda})}=\frac{V_{R}(E',\bm{\lambda}')}{t_{R}(E',\bm{\lambda}')}\\
\frac{T_{R}(E,\bm{\lambda})}{V_{R}(E,\bm{\lambda})}=s\frac{T_{R}(E',\bm{\lambda}')}{t_{R}(E',\bm{\lambda}')}\\
\frac{C_{R}(E,\bm{\lambda})}{V_{R}(E,\bm{\lambda})}=s\frac{C_{R}(E',\bm{\lambda}')}{t_{R}(E',\bm{\lambda}')}
\end{cases}\label{eq:eqs_dualities-1}
\end{equation}
with $s=\pm1$. For SD points, $(E',\bm{\lambda}')\equiv(E,\bm{\lambda})$
and we simply have $V_{R}(E,\bm{\lambda})=\pm t_{R}(E,\bm{\lambda})$
as we have seen before. Note that the freedom in the sign $s$ exists
because we may choose $\varphi_{0}=\pi$ or $\kappa_{0}=\pi$ and
obtain equally valid conditions for the dual points.

Apart from special cases, solving Eqs.$\,$\ref{eq:eqs_dualities-1}
analytically is not possible or impractical \footnote{Note that the solutions for Eqs.$\,$\ref{eq:eqs_dualities-1} that
match true dual points should satisfy $\lim_{E\rightarrow E_{c},\bm{\lambda}\rightarrow\bm{\lambda}_{c}}E'(E,\bm{\lambda})=E_{c}$
and $\lim_{E\rightarrow E_{c},\bm{\lambda}\rightarrow\bm{\lambda}_{c}}\bm{\lambda}'(E,\bm{\lambda})=\bm{\lambda}_{c}$
at self-dual points $(E_{c},\bm{\lambda}_{c})$. There may be other
solutions corresponding to points with equal Fermi surfaces $E(\varphi,\kappa)=E$
at different energy bands. These do not satisfy the previous self-duality
conditions.}. Nonetheless, dual points may still be found numerically similarly
to \citep{HdualitiesScipost}, for a given energy band. 

\section{Reentrant localization transition in PRB 105, L220201 (2022)}

\label{sec:Reentrant-localization-transition}

An interesting reentrant localization transition was observed in \citep{PhysRevB.105.L220201}
for the Aubry-André model with a staggered potential. We here show
that it is possible to capture this transition by inspecting the approximate
self-dual points in CA with unit cells that are small enough to allow
for an analytical calculation.

The Aubry-André model with a staggered potential was also considered
in the Supplementary Materials of \citep{HdualitiesScipost} and it has an
Hamiltonian given by

\begin{equation}
\begin{aligned}H= & -t\sum_{n}(c_{n}^{\dagger}c_{n+1}+{\rm h.c.})+V\sum_{n}\cos(2\pi\tau_{c}n+\phi)c_{n}^{\dagger}c_{n}+\eta\sum_{n}(-1)^{n}c_{n}^{\dagger}c_{n}\end{aligned}
\label{eq:H_staggered}
\end{equation}

This model is particularly challenging because the staggered potential
introduces a relevant perturbation: for finite $\eta$, a CA is only
well defined if the unit cell has an even number of sites, in contrast
with the $\eta=0$ case. For a CA characterized by $\tau_{c}=5/8$,
defining $\varphi=L\phi/2$ (notice the factor of 2 to account for
the periodicity in $\phi$ for finite $\eta$), we have

\begin{equation}
\begin{aligned}\mathcal{P}(\varphi,\kappa)= & t_{R}\cos(\kappa)+V_{R}\cos(\varphi)+V_{2R}\cos(2\varphi)\\
= & 2\cos(\kappa)-\eta E\Big[V^{4}(-2\sqrt{2}+E^{2}+\eta^{2})+\frac{V^{6}}{2}\Big]\cos(\varphi)-\frac{V^{8}}{128}\cos(2\varphi)+\cdots
\end{aligned}
\label{eq:coefs_reentrant}
\end{equation}

Note that the fundamental harmonic in $\varphi$ only arises for $\eta\neq0$
(for $\eta=0$ we get the Aubry-André model and a fundamental harmonic
in $\varphi'=L\phi$). The condition $|t_{R}|=|V_{R}|$ is a good
approximation for the self-duality condition for $\tau$ close to
$\tau_{c}$ as long as $|t_{R}|,|V_{R}|\gg|V_{2R}|$. In Fig.$\,$\ref{eq:coefs_reentrant}
the approximate SD curves obtained through $|t_{R}|=|V_{R}|$ along
with the IPR results for $\tau_{i}=575/918$. The reentrant extended
phase occurs in the region with smaller IPR that arises at larger
$V$, as observed in \citep{PhysRevB.105.L220201}. This reentrant
transition is captured by the SD curves of the CA with $\tau_{c}=5/8$.
The phase transition points do not match exactly the IPR results due
to the approximate nature of these curves. In particular we have that
$|V_{2R}/t_{R}|=|V_{2R}/V_{R}|\approx0.2$ for the values of $V$
around this reentrant transition (the approximation for the smaller
$V$ extended-to-localized transitions is much better as in this case
$|V_{2R}/t_{R}|=|V_{2R}/V_{R}|\approx10^{-5}$). Furthermore, we also
verified that the location of reentrant phase is very sensitive to
$\tau$ and therefore higher-order CAs are needed for more accurate
quantitative phase boundaries, for which an analytical treatment becomes
more challenging. Nonetheless, it is remarkable to see that the reentrant
behaviour is already qualitatively (and almost quantitatively) captured
for a CA with only 8 sites in the unit cell. This happens because
the values of $\eta$ considered here are large enough so that the
original Aubry-André coupling $V_{2R}$ may be seen as a perturbation
compared to the $\eta$-generated coupling $V_{R}$, even for such
a small unit cell.

\begin{figure}[h!]
\centering{}\includegraphics[width=0.4\columnwidth]{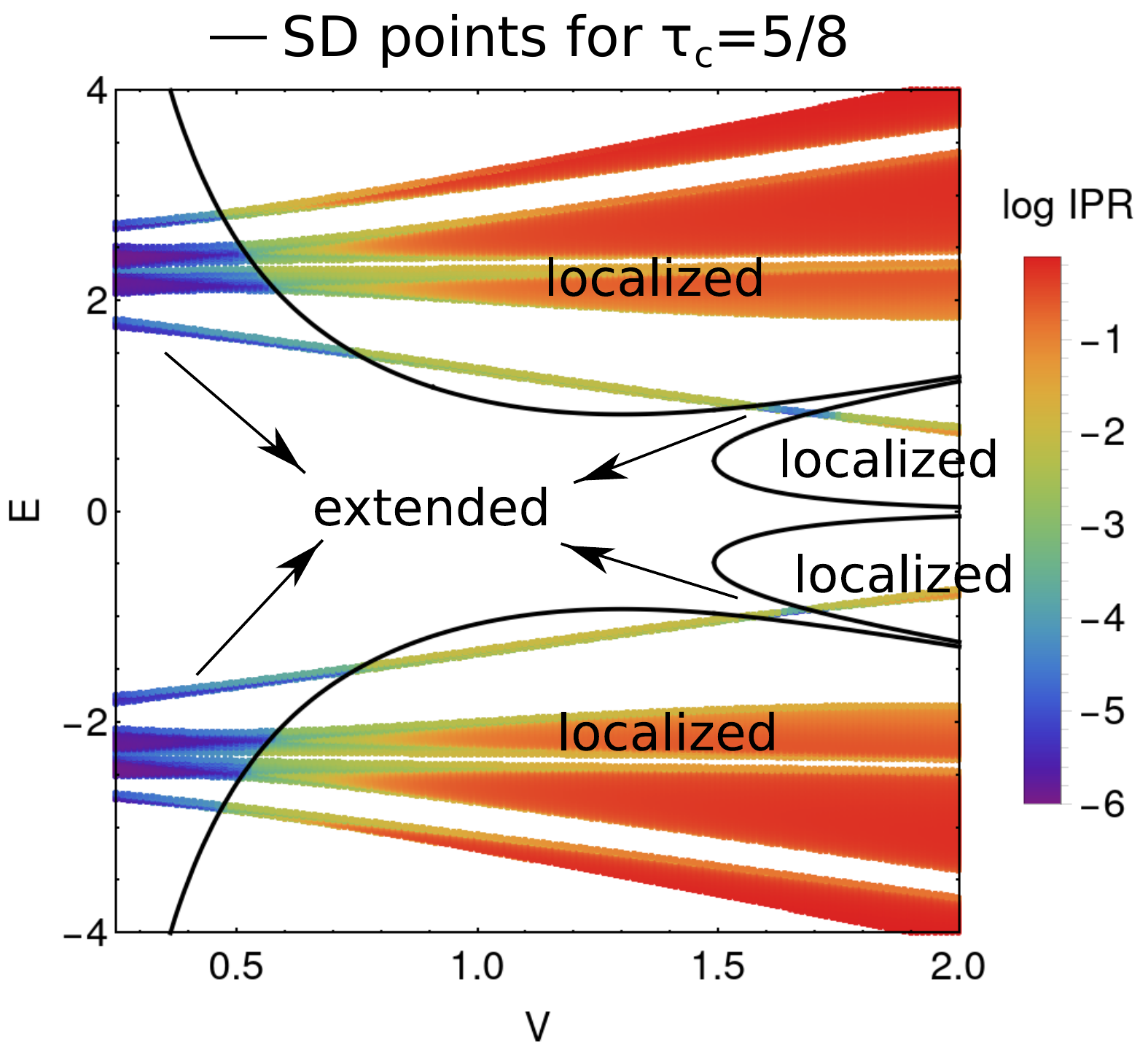}\caption{IPR results for the model in Eq.$\,$\ref{eq:H_staggered} for $\eta=1.8$,
using $\tau_{i}=575/918$, along with the SD curves obtained through
the condition $|t_{R}|=|V_{R}|$ for $\tau_{c}=5/8$ (see Eq.$\,$\ref{eq:coefs_reentrant}),
with the coefficients defined in Eq.$\,$\ref{eq:coefs_reentrant}.}
\end{figure}




\end{document}